# Measuring the bioeconomy economically: exploring the connections between concepts, methods, data, indicators and their limitations

**Sebastián Leavy**

National Institute of Agricultural Technology–INTA, Argentina. Interdisciplinary Center for Research and Studies in Agribusiness–CEPAN, Bioeconomics Research Group, Universidade Federal do Rio Grande do Sul – UFRGS, Brazil; leavy.sebastian@inta.gob.ar

**Gabriela Allegretti**

Universidade de Rio Verde–UniRV. Brazilian Institute of Bioeconomy–INBBIO, Bioeconomics Research Group, Universidade Federal do Rio Grande do Sul – UFRGS, Brazil; allegretti@unirv.edu.br

**Elen Presotto**

Faculty of Agronomy and Veterinary Medicine–FAV. University of Brasília–UnB; elen.presotto@unb.br

**Marco Antonio Montoya**

Faculty of Economics, Management, and Accounting–FEAC, University of Passo Fundo–UPF, Agribusiness Economics and Management Research Group, Brazil; montoya@upf.br

**Edson Talamini***

Department of Economics and International Relations–DERI, Faculty of Economics–FCE, Interdisciplinary Center for Research and Studies in Agribusiness–CEPAN, Bioeconomics Research Group, Universidade Federal do Rio Grande do Sul–UFRGS, Brazil; edson.talamini@ufrgs.br

*Corresponding author: Av. Bento Gonçalves 7712 – Prédio da Agronomia - 1.º Andar - Porto Alegre – RS - Brasil - CEP: 91540-000.
E-mail: edson.talamini@ufrgs.br

# Measuring the bioeconomy economically: exploring the connections between concepts, methods, data, indicators and their limitations

**Abstract**

Despite its relevance, measuring the contributions of the bioeconomy to national economies remains an arduous task that faces limitations. Part of the difficulty is associated with the lack of a clear and widely accepted concept of the bioeconomy and moves on to the connections between methods, data and indicators. The present study aims to define the concepts of bioeconomy and to explore the connections between concepts, methods, data and indicators when measuring the bioeconomy economically, and the limitations involved in this process. The bioeconomy concepts were defined based on a literature review and a content analysis of 84 documents selected through snowballing procedures to find articles measuring 'how big is the bioeconomy?'. The content of the 84 documents was uploaded to the QDA Miner software and coded according to the bioeconomy concept, the methods or models used, the data sources accessed, the indicators calculated, and the limitations reported by the authors. The results of the occurrence and co-occurrence of the codes were extracted and analyzed statistically, indicating that the measurement of bioeconomy (i) need recognize and pursue the proposed concept of holistic bioeconomy; (ii) rarely considered aspects of holistic bioeconomy (3.5%); (iii) is primarily based on the concept of biomass-based bioeconomy–BmBB (94%); (iv) the association with the concept of biosphere–BsBB appeared in 26% of the studies; (v) the biotech-based bioeconomy–BtBB was the least frequent (1.2%); (vi) there is a diversity of methods and models, but the most common are those traditionally used to measure macroeconomic activities, especially input-output models; (vii) depending on the prevailing methods, the data comes from various official statistical databases, such as national accounts and economic activity classification systems; (viii) the most frequently used indicators are value added, employment and GHG emissions; (ix) there are various limitations related to the concept, methods and models, data, indicators and others, like incomplete, missing or lack of data, aggregated data, outdated data or database, and uncertainty of the estimated values, the subjectivity in the bio-shares weighting procedures, and other limitations inherent to methods and models. We conclude that efforts to measure the bioeconomy should be encouraged, starting by recognizing that the measurement of a holistic bioeconomy must be pursued.

**Keywords:** Bioeconomics; Holistic bioeconomy; Bio-based economy; Biotechnology; Biotechonomy; Bioecology; Digitalization.

**List of abbreviations**

A-LCA - Attributional Life Cycle Assessment
ADF - Augmented Dickey-Fuller
AHP - Analytic Hierarchy Process
AIC - Akaike info criterion
ARCH- Autoregressive conditional heteroscedasticity.
ARDL - Autoregressive Distributed Lag
BmBB - Biomass-Based Bioeconomy
BRIC+ - Brazil, China, Indonesia, India, Russia, Taiwan
BsBB - Biosphere-Based Bioeconomy
BtBB - Biotechnology-Based Bioeconomy
BTF - Bioeconomy Transition Framework
BTS - Bartlett's test of sphericity
C-LCA - Consequential Life Cycle Assessment
CAIT - Climate Data Explorer
CAPRI - Common Agricultural Policy Regionalised Impacts
CAWI - Computer-Assisted Web Interview
CBM - Carbon Budget Model
CED - Cumulative Energy Demand
CEE - Central and Eastern European
CES - Constant Elasticity Substitution
CET - Constant Elasticity Transformation
CGE - Computable General Equilibrium Model
CH Rule - Calinski-Harabasz (CH) rule
$CH_4$ - Methane
CICES - Common International Classification of Ecosystem Services
$CO_2$ - Carbon Dioxide
CoCo - Complete and Consistent
COFOG - Classification Of the Functions Of Government
COMEXT - Eurostat reference database for international trade in goods.
CPA - Statistical Classification of Products by Activity in the European Economic Community
CPC - Central Product Classification
CRF - Common Reporting Format
CSO - Central Statistics Office in Ireland
CUSUMQ - Cumulative Sum of Squares
CZSO - Czech Statistical Office
DEU - Domestic Extraction Used
dFRI - Direct Fossil Resource Intensity
DMI - Domestic Material Input
EAA - Economic Accounts for Agriculture
EC - European Commission
ECT - Error Correction Term
EEIO - Environmental Extended Input-Output Model
EEMRIO - Environmental Extended Multi-Regional Input-Output Model

EFISCEN - European Forest Information Scenario Model
EIA - Energy Information Administration
EPA - Environmental Protection Agency
EU - European Union
EU CSO - European Construction Sector Observatory
EU KLEMS - EU level analysis of capital (K), labour (L), energy (E), materials (M) and service (S) inputs
EXIOBASE - Multi-Regional Environmentally Extended Supply-Use Table (MR-SUT) and Input-Output Table (MR-IOT)
FADN - EU Farm Accountancy Data Network
FAO - Food and Agriculture Organization
FAOSTAT - FAO Statistics
FEA - Forestry Economic Accounts
FLQ - Flegg Location Quotient
ISI - Institute for Systems and Innovation Research
FRC - Fossil Resource Consumption
FRS - Fossil Resource Savings
FU - Functional Unit
GAMS - General Algebraic Modeling System
GDP - Gross Domestic Product
GEIH - Gran Encuesta Integrada de Hogares
GFCF - Gross Fixed Capital Formation
GFN - Global Footprint Network
GFTM - Global Forest Trade Model
GHG - Greenhouse Gases
GINFORS - Global Inter-industry Forecasting System
GMFD - Global Material Flows Database
GRIT - Generation of Regional Input–Output Tables
GTAP - Global Trade Analysis Project
GTAP-AEZ - Global Trade Analysis Project Agro-Ecological Zones
GTAP-AGR - Global Trade Analysis Project-Agriculture
GTAP-E - Global Trade Analysis Project-Energy
GTEM - Global Trade and Environment Model
GWP - Global Warming Potential
HAC - Hierarchical Agglomerative Clustering
HDI - Human Development Index
HEM - Hypothetical Extraction Method

ICT - Information and Communication Technology
IEA - International Energy Agency
IMAG - Integrated Model to Assess the Global Environment
IMPACT - International Model for Policy Analysis of Agricultural Commodities and Trade
IMPLAN - Impact Analysis for Planning
IO - Input-Output
IOM - Input-Output Model

IPCC - Intergovernmental Panel on Climate Change
IRENA - International Renewable Energy Agency
ISIC - International Standard Industrial Classification of All Economic Activities
ISTAT – Istituto Nazionale di Statistica
JCR - Join Commission Research
KEI - Knowledge Economic Index
KMO - Kaiser–Meyer–Olkin
LCA - Life Cycle Assessment
LCIA - Life Cycle Impact Assessment
LFS - Labour Force Survey
LPES - Linear Programming Energy System
LQ - Location Quotient
LULC - Land Use and Land Change
MAGNET - Modular Applied General Equilibrium Tool
MARKAL-NL-UU - Market Allocation-The Netherlands-Energy and chemical industry sectors
MCDA - Multi-criteria Decision Analysis
MFA - Material Flow Analysis
MRIO - Multirregional Input-Output
$N_2O$ - Nitrous Oxide
NACE - Statistical Classification of Economic Activities in the European Community
NAFTA - North American Free Trade Agreement
NAICS - North American Industry Classification System
NAMEA - National Accounting Matrices including Environmental Accounts
NASEM - The National Academies of Sciences, Engineering, and Medicine
NEMO - non-European Major OECD countries
NFS - National Farm Survey
OCDE - Organisation for Economic Co-operation and Development
ORBIS - Open Repository Base on International Strategic Studies
PATSTAT - European Patent Database
PCA - Principal Component Analysis
PKD - Polish Classification of Activities
PP - Phillips–Perron Test
PRODCOM - '*PRODuction COMmunautaire*' (Community Production)
QDA - Qualitative Data Analysis
R&D - Research and Development
RAS - Methodology to balance matrices
rDNA - Recombinant Deoxyribonucleic Acid
REGON - National Official Business Register
RESET - Regression Specification Error
ROW - Rest of World
RPA - Revealed Patent Advantage
SAM - Social Accounting Matrices
SAT-BBE - Systems Analysis Tools Framework for the EU Bio-Based Economy Strategy

SBS - Structural Business Statistics or Survey
SEAI - Sustainable Energy Authority of Ireland
SEIB - Socio-Economic Indicator for the Bioeconomy
SIMSIPSAM - Simulation for social indicators and poverty using SAM
SQL - Simple Location Quotient
SSI - Substitution Share Indicator
SSS - State Statistics Service
STECF - Scientific, Technical and Economic Committee for Fisheries
SUTs - Supply and Use Tables
tFRI - Total Fossil Resource Intensity
TOPSIS - Technique for Order Performance by Similarity to Ideal Solution
ToSIA - Tool for Sustainability Impact Assessment
US, BEA - US Department of Commerce's Bureau of Economic Analysis
VEC - Vector Error Correction
WALD Test - Multivariate generalization test developed by Wald (1943)
WIOD - World Input-Output Database
WPINDEX - Derwent World Patents Index

## 1. Introduction

In the last two decades, the bioeconomy has emerged as the promise of a new paradigm toward developing a sustainable economic system (Ferreira et al., 2023). This expectation is based on the renewability of biological resources and the circulation of chemical elements. Thus, the bioeconomic paradigm is superior to the economic system based on finite resources whose stocks are finite and non-renewable (Patermann and Aguilar, 2021). As new political and economic stimuli have been prospected, new ventures have been made in the field of the bioeconomy, leading various countries to adopt specific programs, strategies or guidelines to stimulate the bioeconomy.

The Organisation for Economic Co-operation and Development was the first governmental organization to propose an official agenda for promoting the bioeconomy in 2009 (OECD, 2009). In 2012, the European Union launched its strategy for developing a bioeconomy that connected the economy to nature for sustainable growth (EU 2012, EU 2013). Since then, national governments worldwide have made efforts to promote bioeconomy activities and boost national economies officially. These include the Netherlands, Sweden, the United States and Russia in 2012, Malaysia and South Africa in 2013, Germany and Finland in 2014 and France in 2016 (Priefer et al., 2017). Brazil launched its national bioeconomy program in 2019 (BRASIL, 2019), while China includes the bioeconomy in its 14th five-year plan for 2021-2025 (Zhang et al., 2022). Currently, dozens of countries have specific or related bioeconomy strategies (OECD 2018, Gardossi et al. 2023).

However, one aspect that differentiates national strategies is the essence of the bioeconomy concept considered when outlining the actions to be stimulated. In the EU, for example, the concept was updated in 2018 to add sectors and systems that rely on biological resources, in addition to those primary sectors that produce renewable biological resources (EC 2018, Kardung et al. 2019)[1]. Although it is recognized that there is no widely accepted concept of the bioeconomy (NASEM, 2020), in the United States, it can be said that former President Barack Obama's National Bioeconomy Blueprint is based on three main and correlated axes: knowledge, technology and innovation (US, 2012)[2]. From the North American perspective, the bioeconomy is the economic activities driven by research and innovation in the life sciences and biotechnology, and that are enabled by technological advances in engineering and computing and information sciences (NASEM 2020, Frisvold et al. 2021). Genetic engineering, molecular biology, bioinformatics, and synthetic biology are some driving technological platforms. The Bioeconomy Brazil Program focuses on extracting renewable resources and exploring biodiversity as a strategy for the sustainable development of smallholder and indigenous communities (BRASIL, 2019)[3]. Although other countries adopt slightly different concepts (see Box 2.1 in NASEM 2020), the concepts adopted by the EU, US and Brazil are examples that align with the visions of the bioeconomy reported in scientific literature.

---

[1] "The bioeconomy covers all sectors and systems that rely on biological resources (animals, plants, micro-organisms and derived biomass, including organic waste), their functions and principles. It includes and interlinks: land and marine ecosystems and the services they provide; all primary production sectors that use and produce biological resources (agriculture, forestry, fisheries and aquaculture); and all economic and industrial sectors that use biological resources and processes to produce food, feed, bio-based products, energy and services" (EC 2018).

[2] "A critical factor in leveraging biological systems to drive an innovation-based bioeconomy is the strength of the scientific enterprise investigating those systems, including basic and applied research. A robust biological/biomedical R&D enterprise, backed by government, foundations, and for-profit investments, is necessary to produce the new knowledge, ideas, and foundational technologies required to develop products and services that support businesses and industries and help create jobs. Nature has evolved countless biologically based systems with potential for new applications to address problems in health, energy, food, and the environment. Expanding basic knowledge of living systems and their molecular machinery will inspire new concepts for the creation of artificial processes and products that will address current and future needs. A sustained effort to understand and take advantage of natural living systems will produce novel solutions and encourage the growth of the bioeconomy" (US, 2012).

In scholarly circles, several authors have differentiated bioeconomy approaches by considering the resources, processes, and effects that activities trigger in the economic and biophysical systems. Bugge, Hansen, and Klitkou (2016) classified the bioeconomy into (i) bioresources, indicating products composed wholly or in large part of biomass; (ii) biotechnology, products or processes enabled by innovation in the life sciences; and (iii) bioecology, based on sustainable processes and sustainably sourced products. Vivien et al. (2019) categorized the bioeconomy into (i) Type I, an ecological economy, that is compatible with the biosphere constraints; (ii) Type II, a science-based economy driven by industrial biotechnology; and (iii) Type III, a biomass-based economy. Befort (2020) categorized the visions of the bioeconomy based on socio-technical regimes, proposing (i) Biotech-bioeconomy, which is technology-driven, spreading biotechnology into health, food, industry, etc., and industrialization of the living; and (ii) Biomass-bioeconomy, whose mission-driven towards the substitution of oil for biomass. Other variants of classifications, typologies and narratives are explored in the literature, as in Wei, Liu, et al. (2022) and Allain et al. (2022). However, the classifications cited are sufficient to (i) show evidence that a widely accepted concept of the bioeconomy is lacking; (ii) link the national concepts to the visions of the bioeconomy in the literature, being EU closed to bioresources or Type III, US to biotechnology or Type II, and Brazil to bioecology or Type I; and (iii) reflect on what are the implications for the measurement of the bioeconomy.

[3] "The *Programa Bioecomia Brasil (Sociobiodiversidade)* aims to promote partnerships between public authorities, small farmers, family farmers, traditional peoples and communities and their enterprises, and the business sector, to promote and structure production systems based on the sustainable use of sociobiodiversity resources and extractivism, as well as the production and use of energy from renewable sources that allow these segments to participate more in production and economic arrangements that involve the concept of the bioeconomy" (BRASIL, 2019).

Despite these conceptual differences, there is interest and effort on the part of stakeholders to show the economic importance of the bioeconomy and to monitor indicators that attract investment to bioeconomy-related activities. Although awareness differ between stakeholders, the economic dimension is still relevant (Zeug et al., 2021). In this regard, Tévécia Ronzon et al. (2020) estimated that the bioeconomy contributed 4.7% of the EU-27's GDP in 2017. In the US, the contribution of the biobased bioeconomy to the national GDP in 2016 was 5.1% and could reach 7.4% (NASEM, 2020). Specifically, the biotechnology-based bioeconomy would have reached 2.0% of the US GDP in 2012, up from 0.004% in 1980. An annual growth rate of over 10% (Carlson, 2016). In Brazil, the bioeconomy value chain accounted for 19.6% of GDP in 2019 (Cicero Zanetti Lima and Pinto, 2022), a higher share than the 13.8% of added value found by Silva, Pereira, and Martins (2018).

As highlighted above, there are different concepts and values for the contribution of the bioeconomy to national economies. Monitoring schemes and indicators are currently under development to assess the impacts of bioeconomy and better inform policy-making, but it can be considered a Herculean task (Jander et al. 2020). One question that matters is: how concepts, methods, data and indicators are connected in the process of measuring bioeconomy economically, and what are the limitations involved? A relevant study in this regard is the review by Ferreira et al. (2022) which lists methods, variables and data used to measure the sustainability of the bioeconomy, including the economic dimension. However, the study by Ferreira et al. (2022) does not define concepts nor explore the connections between methods, data, and indicators and their limitations. The present study advances in detailing these connections when applied in the context of measuring 'how big is the bioeconomy?' (Kuosmanen et al., 2020a). Therefore, the present study aims to define the concepts of bioeconomy and to explore the connections between concepts, methods, data and indicators when measuring the bioeconomy economically, and the limitations involved in this process.

## 2. Materials and Methods

The first step to reach the goal of exploring the connections between concepts, methods, data and indicators when measuring the bioeconomy economically, and the limitations involved in this process was to define the concepts of bioeconomy. Once the concepts of the bioeconomy have been defined, the next step was to carry out a content analysis of carefully selected documents. This section describes the methodological procedures used to complete these research steps. The purpose of providing these details is two-fold: First, to inform the reader about how the results were obtained, and second, to ensure that other researchers can replicate the study. The following steps to achieve these goals are explained in detail.

### *2.1. First Step: Defining the concepts of bioeconomy*

As mentioned earlier, there is no widely accepted definition for the bioeconomy that incorporates all its idiosyncrasies. However, defining the concept of bioeconomy is crucial for exploring its connections with methods, data, and indicators. To achieve this, a literature review was conducted on scientific articles that reported conceptual analyses, visions, narratives, or approaches to the bioeconomy. The concepts were then organized based on the similarity of their fundamental elements. Three concepts were defined as partial, as they adopted a restricted vision of bioeconomy. Based on the characteristics and limitations of these partial concepts, a concept of holistic bioeconomy was proposed that combines many characteristics and elements of partial concepts. This new concept has advantages for measuring the true value of the bioeconomy.

### *2.2. Second Step: Articles selection and content analysis*

This section describes the procedures adopted for searching and selecting articles, building the database, defining the codes, coding the articles' content and extracting the results.

a. *Searching for articles*: content analysis requires selecting documents of interest. In this respect, the scope of documents to be retrieved was restricted to measuring bioeconomy economically. The methodological strategy chosen to find documents (scientific articles, technical reports, government reports) fitting to the desired scope follows the snowballing procedures.

Snowballing is a well-established and widely used technique for exploring related documents by identifying articles relevant to your topic of interest from key and cited-by articles (Greenhalgh and Peacock, 2005). In some cases, snowballing proved to be a more efficient method for retrieving articles of interest than the databases themselves (Badampudi et al., 2015), depending on the database (Felizardo et al., 2018). Synthetically, according to Greenhalgh and Peacock (2005), snowballing refers to using the reference list of a paper or the citations to the paper to identify additional papers. Start with a few articles currently in or around your topic of interest. These articles are referred to as 'start set.' Once the start set is defined, it is time to conduct backward and forward snowballing.

The snowballing procedures followed the steps described by Wohlin (2014), Wohlin (2016) and Wohlin et al. (2022). According to Jalali and Wohlin (2012), systematic literature studies can be done differently regarding the first step. Based on the specificity of the study, we opted to start with the reference lists of a starting set of papers, called the 'start set.' The start set of papers was identified in two previous studies by Wesseler and von Braun (2017) and Cingiz et al. (2021), where a list of papers about measuring bioeconomy was presented in Tables 2 and 1, respectively. By merging both tables and excluding duplicated papers, a list of 10 papers remains as the start set, as listed in Table 1.

**Table 1.** List of papers serving as the start set for snowballing procedures.

| **Document order and information** | **Reference** |
|---|---|
| 1. Nowicki P, Banse M, Bolck C, Bos H, Scott E. 2008. Biobased economy. State-of-the-art assessment. Rep., Agric. Econ. Res. Inst., The Hague, The Netherlands. | (Nowicki et al., 2008) |
| 2. Vandermeulen V, Prins W, Nolte S, Van Huylenbroeck G. 2011. How to measure the size of a bio-based economy: evidence from Flanders. Biomass Bioenerg. 35: 4368–75. | (Vandermeulen et al., 2011) |
| 3. Efken J, Banse M, Rothe A, Dieter M, Dirksmeyer W, et al. 2012. Volkswirtschaftliche Bedeutung der biobasierten Wirtschaft in Deutschland. Work. Pap., 07/2012, Johann Heinrich v. Thünen-Inst., Braunschweig, Germany.(*) | (Efken et al., 2012) |
| 4. Rosegrant MW, Ringler C, Zhu T, Tokgoz S, Bhandary P. 2013. Water and food in the bioeconomy: challenges and opportunities for development. Agric. Econ. 44(s1):139–50. | (Rosegrant et al., 2013) |
| 5. Golden JS,Handfield RB, Daystar J, McConnell TE. 2015. An economic impact analysis of the U.S. biobased products industry: a report to the Congress of the United States of America. Ind. Biotech. 11(4):201–9. | (Golden et al., 2015) |
| 6. Ronzon T, Santini F, M'Barek R. 2015. The bioeconomy in the European Union in numbers. Facts and figures on biomass, turnover and employment. Rep., Eur. Comm., Joint Res. Cent., Inst. Prospect. Tech. Stud., Sevilla, Spain. | (Ronzon et al., 2015) |
| 7. Heijman W. 2016. How big is the bio-business? Notes on measuring the size of the Dutch bio-economy. NJAS 77:5–8. | (Heijman, 2016) |
| 8. Carlson R. 2016. Estimating the biotech sector's contribution to the US economy. Nat. Biotechnol. 34(3):247–55. | (Carlson, 2016) |
| 9. Ronzon, T.; Piotrowski, S.; Tamosiunas, S.; Dammer, L.; Carus, M.; M'barek, R. 2020. Developments of Economic Growth and Employment in Bioeconomy Sectors across the EU. Sustainability, 12, 4507. | (Ronzon et al., 2020) |
| 10. Kuosmanen, T.; Kuosmanen, N.; El Meligi, A.; Ronzon, T.; Gurria Albusac, P.; Iost, S.; M'Barek, R. 2020. How Big is the Bioeconomy? Reflections from an economic perspective; Publications Office of the European Union: Luxembourg. | (Kuosmanen et al., 2020b) |

Note: documents ordered by publication date. (*) Efken et al. 2012 was included in the start set papers but was excluded of content analysis because it is in German language.
Source: papers selected from Wesseler and von Braun (2017) and Cingiz et al. (2021).

We began the backward and forward snowballing procedures from the start set papers. Backward snowballing means using the reference list of a starting set paper to identify new documents to include in the literature review, while forward snowballing refers to identifying new papers based on those papers citing the paper being examined (Wohlin 2014). A series of iterations were conducted during backward and forward snowballing. The following procedures were carried out to accomplish the first iteration. Firstly, we took each of the papers from the start set and wholly revised the list of references, searching for potentially additional papers to be included in the review database. The question 'How big is the bioeconomy?' (Kuosmanen et al., 2020a) illustrates the guiding question we sought in the papers. Secondly, we went to the Scopus database, searched for each paper, and accessed the papers citing it. In the case documents like reports were not indexed in Scopus, we investigated for their citation by filing the query within 'references' with the document title. In both the backward and forward snowballing process, the title of the cited or citing papers was evaluated, and, according to the adherence to the goal of the present study, the papers were pre-selected to compose the review database. Then, the pre-selected papers were revised according to the criteria described in the item b - Selecting the articles. From the selected documents, the second iteration started, and the backward (cited) and forward (citing) snowballing was carried out. This process was repeated in five iterations until no additional papers were selected, as shown in Figure 1.

b. *Selecting the articles*: the criteria to include an article in the database are as follows: (i) the word 'bioeconomy' or 'bio-economy' should appear in the title, abstract, keywords, or at least in the introductory contextualization; (ii) the study should measure some economic aspects of a sector or a bioeconomy system within a local, regional, national, or multi countries' economies; and, (iii) the study should describe a method to measure the bioeconomy share in the economic system. Studies based on reviews, analyzing national strategies or policies on bioeconomy, describing the bioeconomy based purely on official statistical data or conceptual frameworks, or assessing only socio-environmental aspects of bioeconomy sustainability were excluded from the analysis.

At the end of the snowballing process, 84 documents, including articles and technical or scientific reports, were selected to have their content analyzed. The list with 84 selected documents can be accessed in Appendix A (Supplementary Material). The following sections describe the procedures related to the content analysis.

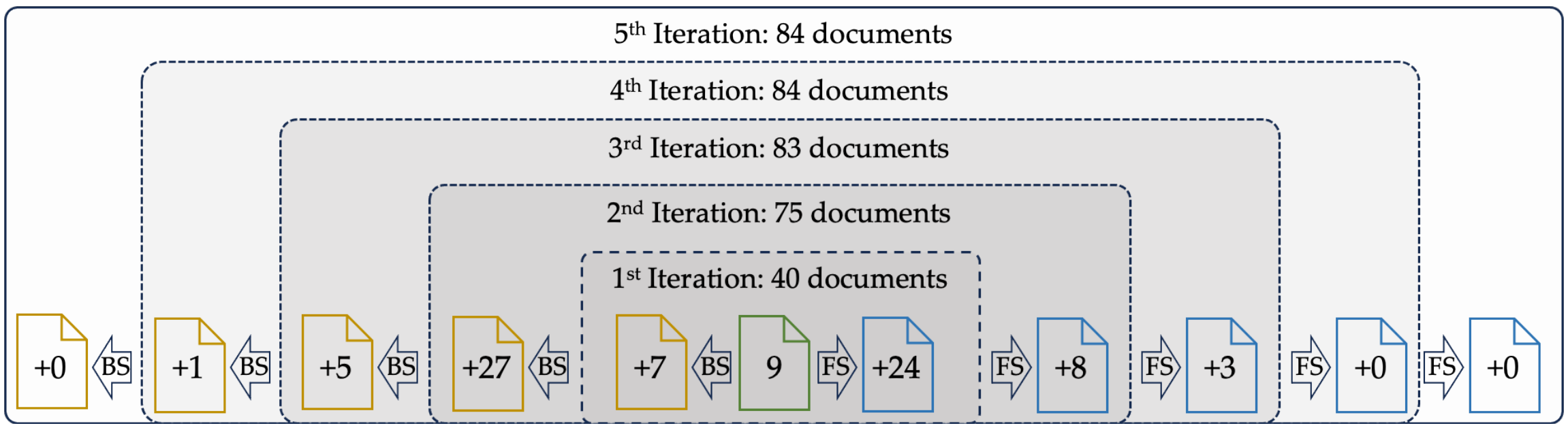

**Figure 1**. Documents identified and selected during the backward (BS) and forward (FS) snowballing iterations from the seed references.
Source: elaborated by the authors based on Wohlin et al. (2022).

c. *Building the database*: The first step to analyzing the documents' content was the construction of an appropriate database. To make the process more reliable, we used the software Qualitative Data Analysis–QDA Miner (v. 5), provided by Provalis Research (https://provalisresearch.com/products/qualitative-data-analysis-software/). The textual content of each document was uploaded to the QDA environment with their year of publication and first author name being assigned as related variables. The associated variables are interesting for comparing results over time and by authors. We opted to compare authors and studies in the present study and did not conduct a temporal analysis. At the end of this process, the textual content of the documents was electronically available to be coded in the QDA Miner environment using the software's functions.

d. *Defining the codes*: We did not use automated text mining processes in this study. Instead, the content of the documents was personally coded by the authors. The textual content of the documents can be coded using a list of predefined codes or an appropriately designed code structure according to the purpose of the study. Although a list of possible and useful codes could be extracted from literature in studies like Wesseler and von Braun (2017), Karvonen et al. (2017), Lier et al. (2018), Giuntoli et al. (2020), Cingiz et al. (2021) and Ferreira et al. (2022), we decided to construct a particular code structure.

Considering the purpose of the study, we defined a fundamental root-code structure based on concepts, methods, data, indicators and limitations. The studies' geographical scope (country, region, etc.) was also considered a root code. Only the codes under the root code 'bioeconomy concepts' were predefined based on Sections 2.1 and 4.1. of this article. Thus, only the codes BmBB, BtBB and BsBB could be used in the coding process. The codes for the other root codes were defined as the content was analyzed and coded. In this sense, as new codes were identified, they were added to the code structure. As the list of codes expanded, the codes began to repeat themselves, and the occurrence of additional codes ceased. The final code structure can be found in Appendix B (Supplementary Material).

e. *Coding the article's content*: With the textual database and code structure available in the QDA Miner user interface, the next step was to code the content of the documents. The coding process involves linking a given code to a fraction of the corresponding text. For example, if, during the content analysis, a passage of text that characterized the use of the Biomass-based Bioeconomy concept was identified, this text fragment was selected, and the code 'BmBB' was linked to it. The same procedure was applied to the other codes relating to methods, data, indicators and limitations. Therefore, at the end of the coding process, the content of the 84 documents was coded entirely in terms of aspects relating to bioeconomy concepts, measurement methods, data used, indicators generated and any limitations identified. Finally, a complete review of duplicate codes in the same document was completed, and corrections were made where necessary.

f. *Extracting the results*: A series of results can be extracted from the coded content. In particular, our interest was exploring results that could satisfactorily achieve the study’s goal. In this sense, QDA Miner's statistical resources made it possible to produce some figures directly, and the occurrence and co-occurrence of codes were downloaded so that other figures could be produced in other software.

## 4. Results

### *4.1. The concepts of bioeconomy: from partial to holistic bioeconomy*

The lack of a widely accepted definition of the bioeconomy has made it difficult to compare the contributions of the bioeconomy between countries, as well as to plan public policies and make decisions in sectors and activities related to the bioeconomy (Bracco et al. 2018, Highfill and Chambers 2023). In this section, we propose the concept of holistic bioeconomy based on three partial conceptual approaches discussed in the literature and the recognition that these concepts are not mutually exclusive. We argue that the holistic bioeconomy should be considered and pursued in measuring the bioeconomy.

#### *4.1.1. Biomass-based Bioeconomy–BmBB*

In line with the work of Bugge, Hansen, and Klitkou (2016), Vivien et al. (2019), Befort (2020), and Giampietro (2019), this approach seeks to measure the bioeconomy from a concept that bases the bioeconomy on the primary input, biomass or bioresources. Biomass is any primary or secondary organic matter (waste) derived from plants, animals or microorganisms that can be used as a renewable input to produce energy, food, fibers or materials (IEA, 2022). Measuring BmBB, therefore, starts by identifying the primary sectors that produce biomass, which could suggest a narrow view of the bioeconomy. Studies such as that by Tévécia Ronzon, Iost, and Philippidis (2022a), have identified agriculture, livestock and forestry as the primary sectors of the BmBB. However, methodological approaches may differ regarding the method, data and indicators used to measure the BmBB. The fact is that even though biomass is essentially produced in three or four economic sectors, it flows between almost all sectors of the economic system, being consumed or transformed with added value (Figure 2). Thus, measuring

BmBB provides information on physical or monetary flows that express how large BmBB is about an economy's GDP, for example.

### *4.1.2. Biotechnology-based Bioeconomy–BtBB*

BtBB is the nomination given to the second approach to measuring the bioeconomy. Alternatively, the term 'biotechonomy' can be applied to define the scope of this approach (Blumberga et al., 2016). Conceptually, this approach aligns with the view of bioeconomy as synonymous with biotechnology suggested by Bugge, Hansen, and Klitkou (2016), Wei et al. (2022) and Wei, Luo, et al. (2022) and the Type II bioeconomy identified by Vivien et al. (2019) and reinforced by Befort (2020). This approach is based on biotechnologies developed from biological assets (biodiversity) and used to solve society's problems on different fronts: human health and medicine (Red), agriculture (Green), marine (Blue) industry (White), food and nutrition (Yellow), deserts and arid regions (Brown), bioterrorism and biocrime (Black), intellectual properties, patents and publications (Violet), nanobiotechnology and bioinformatics (Gold) environmental protection (Grey) (see Table 1 in Wei, Liu, et al. 2022). BtBB has gained strength from advances in genomics, synthetic biology, genetic engineering, and other modern biotechnology platforms (Kircher et al., 2022). Therefore, while BmBB is based on raw materials, BtBB is based on knowledge, innovation and processes for adding value to biodiversity, including biomass. Measuring BtBB involves identifying the relevant activities and sectors related to biotechnology. Resources, processes and products are integrated into the dynamics of the economic system, increasing the added value and contribution of BtBB in national economies. The intermediate part of Figure 2 represents this conceptual approach. Studies on BtBB measurements suggest investment in R&D and the number of biotechnology patents as indicators (OECD 2023; Highfill and Chambers 2023).

### *4.1.3. Biosphere-based Bioeconomy–BsBB*

The third conceptual approach to the bioeconomy is the Biosphere-based Bioeconomy (BsBB). While BmBB and BtBB are closely associated with inputs and the technological means for adding value, respectively, BsBB is the broadest and most integrative approach of all. BsBB meets the scope of the bio-ecological vision of the bioeconomy by Bugge, Hansen, and Klitkou (2016) and the Type I Bioeconomy proposed by Vivien et al. (2019), which associates economic activity with the limits of the biosphere. Therefore, while the BmBB and BtBB approaches consider the national economy as a closed system (open only to the economic system of other countries), the BsBB necessarily implies considering the economic system as an open system that interacts with the biophysical system (biosphere). The biosphere provides the necessary substrates for maintaining and developing all biodiversity, which is the basis of the holistic bioeconomy.

### *4.1.4. Holistic Bioeconomy*

The BsBB perspective is relevant because it makes it possible to measure the flows from the biosphere into the economic system and from the economic system into the biosphere. The BtBB intermediates some of these flows, while others are directly exchanged between the BsBB and the BmBB. Therefore, the consumption of flows and stocks and the role of funds for the absorption of waste and pollution become measurable, and aspects relating to the sustainability of the bioeconomy can be gauged (Dafermos, Nikolaidi, and Galanis 2017, Giampietro 2019).

In the 1990s, Odum proposed a challenging task aligning ecology and economics through a new method–the emergy analysis. He aimed to assess the value of energy and materials involved in biological or economic processes converting them into a common unit (sej-solar emjoules). The emergy accounts for directly and indirectly solar energy required to produce the services and products of ecosystems and the national economies (Odum, 1996). It allows calculating the ratio of emergy and money, which is called "emdolar", and reveals how much emergy is needed to produce U$ 1 GDP. However, the complexity of this methodology and the lack of an interdisciplinary approach have limited its application in economic science.

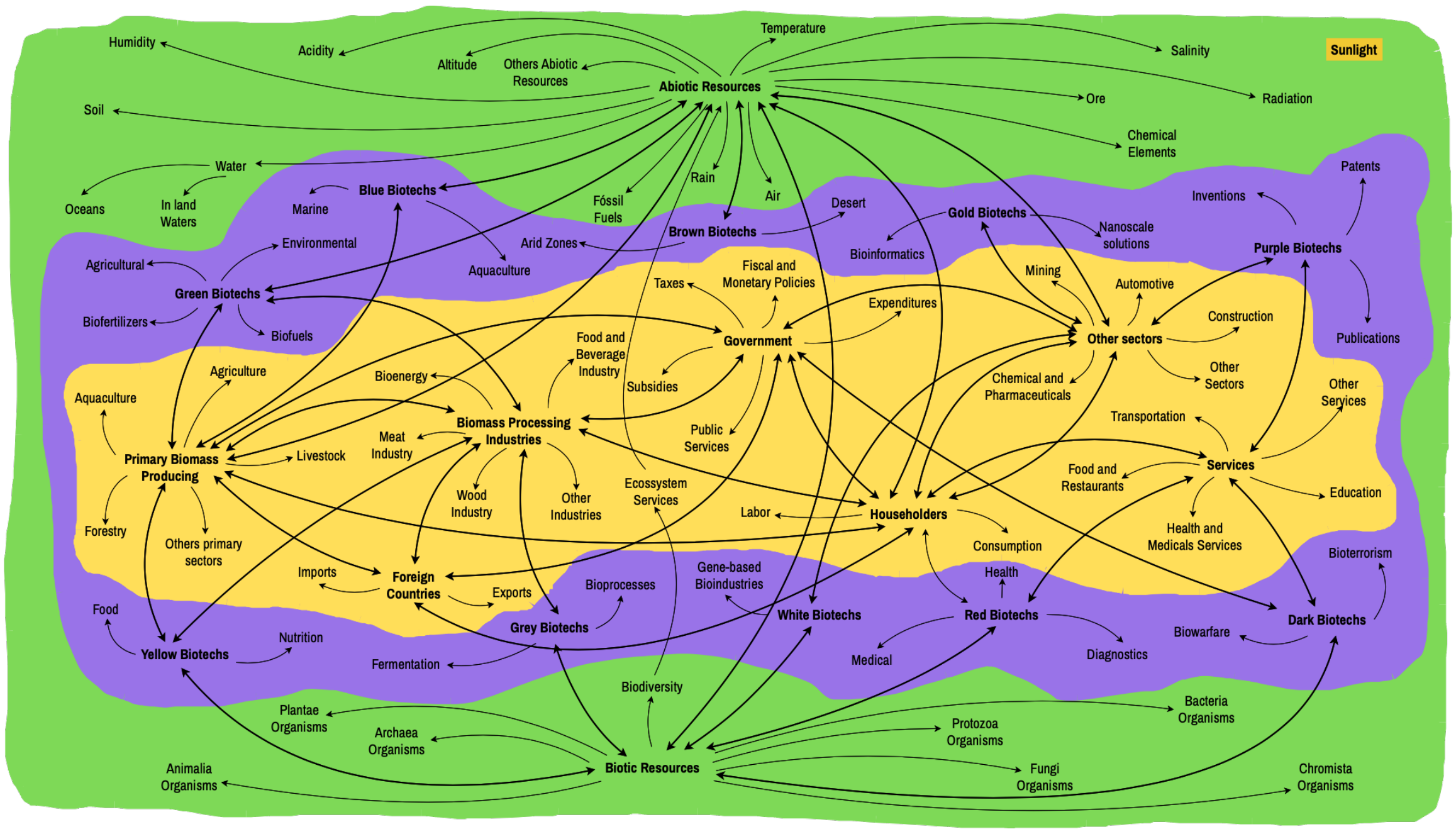

**Figure 2**. The conceptual framework of holistic bioeconomy. In general terms, the central part of the figure (orange) represents the concept of BmBB, indicating the production of biomass as a fundamental input and its intermediate and final flows between the intermediate demand of the economic sectors and the final demand of the government, households and the external market. The intermediate part (purple) represents the concept of BtBB in its different categories of development of biotechnological solutions. BtBB interfaces with the economic system to which it provides its solutions and receives the knowledge of human capital to apply to the biotic resources of the different kingdoms of nature (BsBB). The outermost part of the figure represents the concept of BsBB (green) or, more specifically, the biosphere where the biotic and abiotic resources are located. The biosphere provides the necessary substrates for maintaining and developing all biodiversity, which underpins the holistic bioeconomy.
Source: elaborated by the authors.

Despite this, the BsBB approach cannot be ignored in the methods developed for measuring national bioeconomies. The bioeconomy's contributions to sustainability and the environment have already been highlighted in studies such as Bracco et al. (2018) and Highfill and Chambers (2023). Vivien et al. (2019) suggest that lack of emphasis on BsBB is due to a hijacking of Georgescu-Roegen's term 'Bioeconomy' by economic activities that want to convey a sustainability appeal.

On the other hand, the intransigence of the laws of thermodynamics on which the BsBB is based makes this approach challenging to measure. The laws of thermodynamics may play the role of 'uncomfortable knowledge' in assessing the economy-biophysical interface (Giampietro and Funtowicz 2020, Eversberg et al. 2023). However, if we want to take the bioeconomy seriously in its entirety, we need to focus on developing suitable methods to measure its dimension and its implications. Far from a disciplinary approach grounded in economics, an interdisciplinary engagement grounded in a post-normal science will be decisive (Giampietro and Funtowicz, 2020).

Therefore, the holistic bioeconomy is a comprehensive approach that involves assessing all relationships between BsBB, BtBB and BmBB. The central issue is that the elements of each concept that contribute to the holistic bioeconomy are expressed in different units of measurement. Direct economic measurement of the holistic bioeconomy is unfeasible because their relationships are not always measured in economic terms (monetary value). Therefore, the system represented by the holistic bioeconomy is complex and challenging to measure, but it needs to be considered, as explained by Giampietro (2019) and Giampietro (2024).

*4.2. Authors, documents and countries: A brief characterization*

This section characterizes the set of documents regarding their content similarity and shows the countries leading the studies on measuring the economic impact of bioeconomy. Such results are essential to understand the context from which the main results originated.

The first set of results shown in Figure 3 illustrates clusters of documents grouped by degree of similarity. Cosine similarity is one of the most popular similarity measures applied to text documents, such as in numerous information retrieval applications and document clustering (Huang, 2008). Cosine similarity is a measure of similarity between two vectors obtained from the cosine angle multiplication value of two vectors being compared (Lahitani et al., 2016). Although the Cosine similarity index can take values between -1 and 1, the values range between 0 and 1 in this study. A zero value indicates total dissimilarity between the documents, while a value of 1 means that the documents are entirely similar. Documents with a Cosine index greater than 0.5 can be considered to have high similarity. Although Cosine similarity is widely applied in document content analysis, in the case of this study Cosine similarity is based on the occurrence of codes. Therefore, documents with the same codes would result in a cluster with a Cosine similarity value 1.

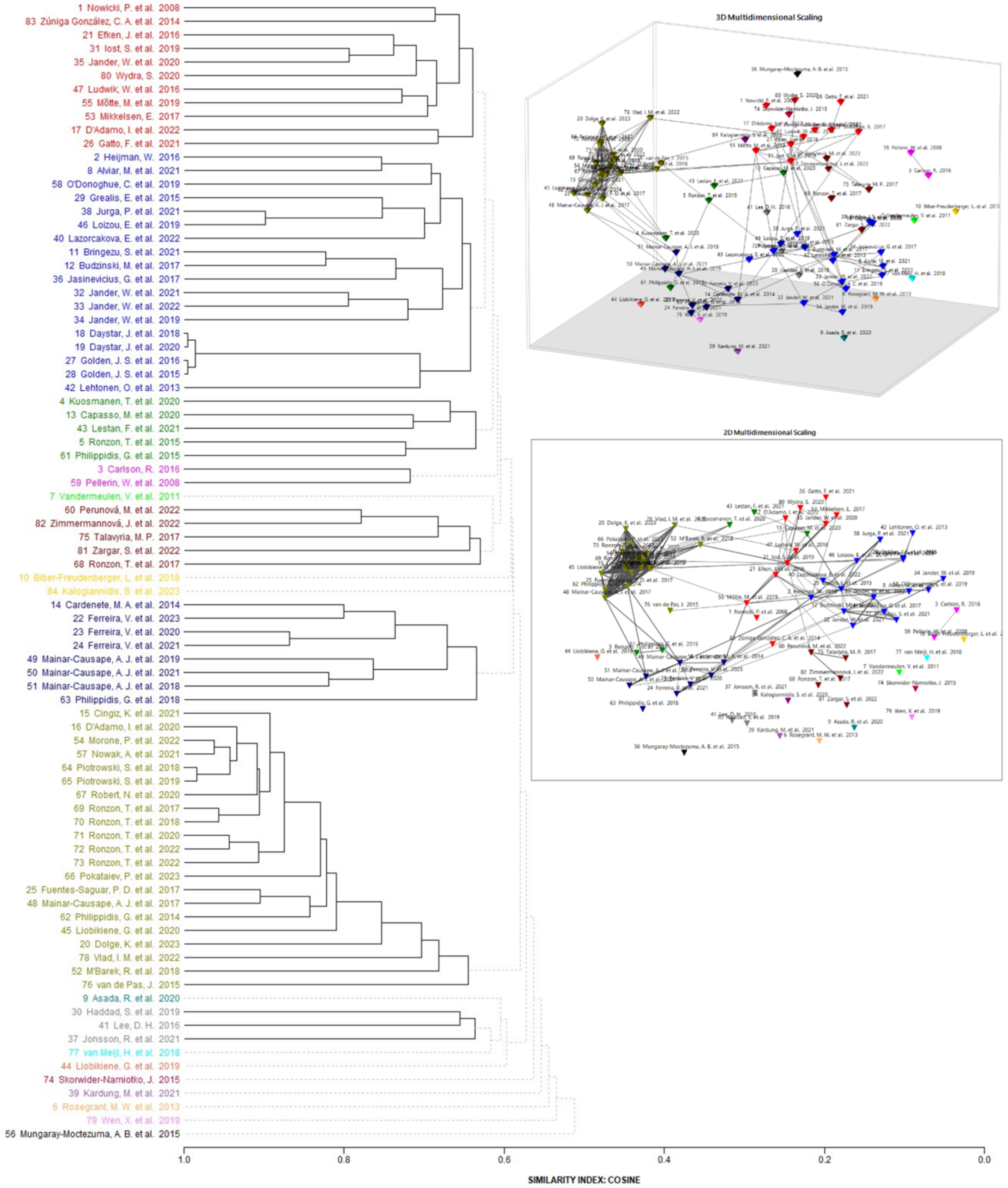

**Figure 3**. Dendrogram, 3D, and 2D multi-dimension scaling of documents clustered by Cosine Similarity Index.

Source: elaborated by the authors based on research data.

Ten main document clusters can be considered based on the similarity of code occurrence. Overall, there is a high degree of similarity between the 84 documents, given that the lowest Cosine index value was higher than 0.5 for all documents. In particular, some documents showed very high similarity, with a Cosine index close to 1. This is the case, for example, with the documents '18 Daystar, J. et al. 2018', '19 Daystar, J. et al. 2019', '27 Golden, J. S. et al., 2016' and '28 Golden, J. S. et al., 2015'. This group of documents consists of a series of technical reports drawn up by a team of researchers and addressed to the US Congress. These authors co-authored the four documents, and each year, the report updates the data using basically the same method, data, and indicators. Another example is documents 64 and 65 by Piotrowski et al. Therefore, the high level of similarity is due to the recurrence of the same codes.

Another feature to be highlighted in the formation of clusters is the effect of co-authorship on the grouping of documents. Many of the authors who form the clusters have worked together with the authors of other documents. In this sense, the cluster formed by the documents '15 Cingiz, K. et al. 2021' to '76 van de Pas, J. 2015' is worth highlighting. These clusters represent the efforts of research groups developing joint studies to measure the bioeconomy using relatively similar concepts, methods, data and indicators.

The high level of similarity between the documents may also have been affected by the geographical scope of the studies. Only 47 countries had data and indicators from their national economies investigated. However, most of the documents report bioeconomy results from countries that are members of the European Union. Germany has the highest number of records, followed by Poland, Spain, the Netherlands and the Czech Republic (Figure 4). Some documents are specific to one country, while others report results from groups of countries, such as EU-28, EU-27, EU-15, and so on. In these cases, the country code was assigned to the document whenever country-specific results could be identified.

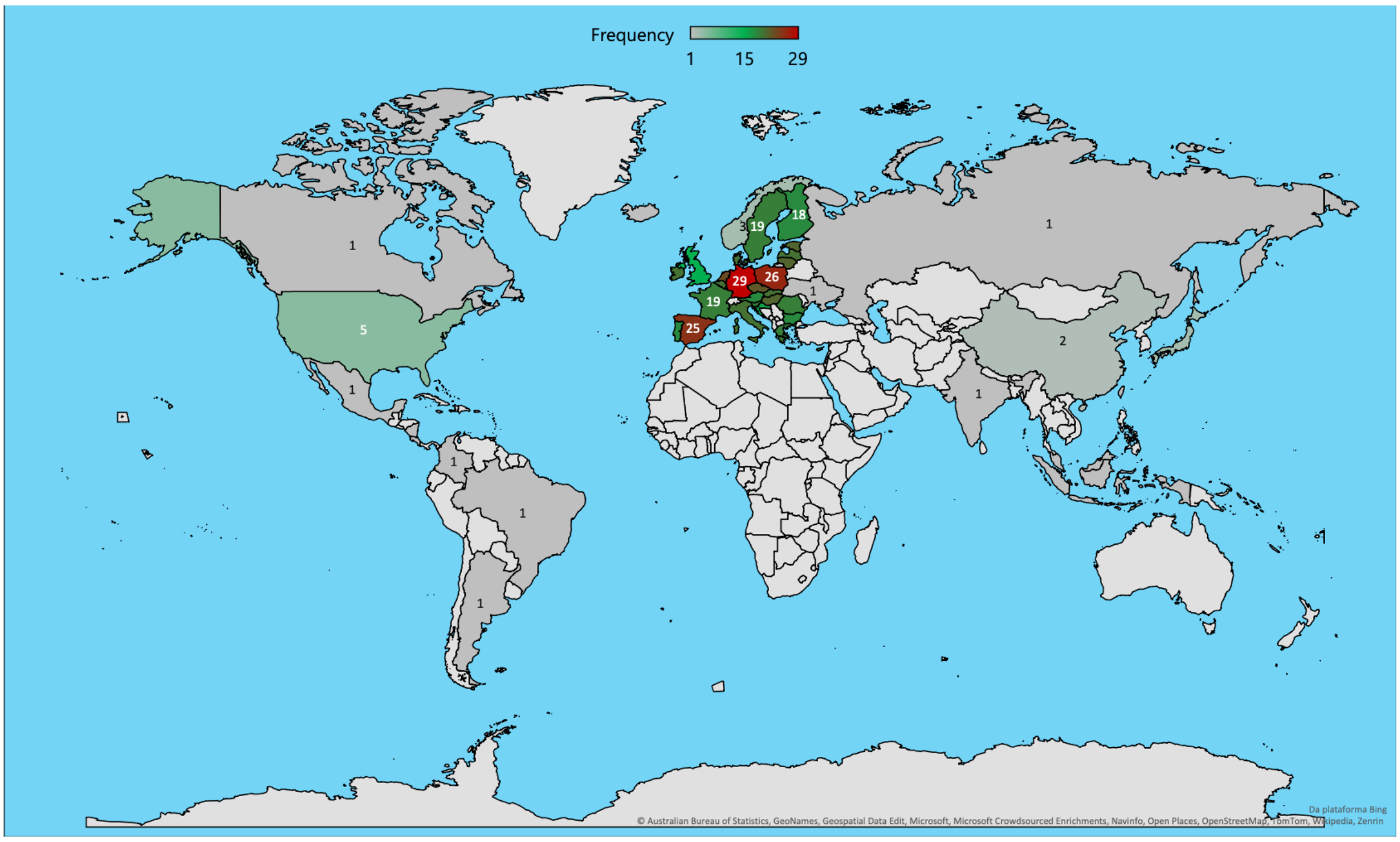

**Figure 4.** Frequency of studies by country.
Source: elaborated by the authors based on research data.

Outside the EU, only some countries have been included in studies. Of these, the United States had the highest number of studies on its bioeconomy (5), followed by Japan (3), Norway (3), and China (2). Despite the potential of the bioeconomy, countries in Latin America, Africa, Oceania and Asia need studies that measure their bioeconomy.

These initial results illustrate a specific pattern in efforts to measure the bioeconomy. Although the studies used various methods, data, and indicators, studies that proposed cutting-edge or significantly innovative methodologies still needed to be identified.

### *4.3. Connections between concepts, methods, data, indicators and limitations*

This section is dedicated to presenting the results relating to the connections between bioeconomy concepts, the methods applied to measure them, the data used to calculate indicators, the indicators measured and the limitations related to these aspects. Figure 5 gives an overview of these connections, showing the most frequent concepts, methods, data, indicators and limitations.

At this point, a result worth highlighting is that most studies (94%) measured the bioeconomy associated exclusively with biomass-based bioeconomy (BmBB). This is likely due to the concentration of studies in EU member countries, where the official concept of the bioeconomy is more aligned with the biobased economy. In 22 documents, economic and environmental aspects were measured together, and these documents were coded with both concepts (BmBB + BsBB). Three documents measured aspects of the three bioeconomy concepts (BmBB + BsBB + BtBB). Only Carlson's (2016) work was dedicated to measuring aspects of Biotechnology-based Bioeconomy (BtBB) in the United States.

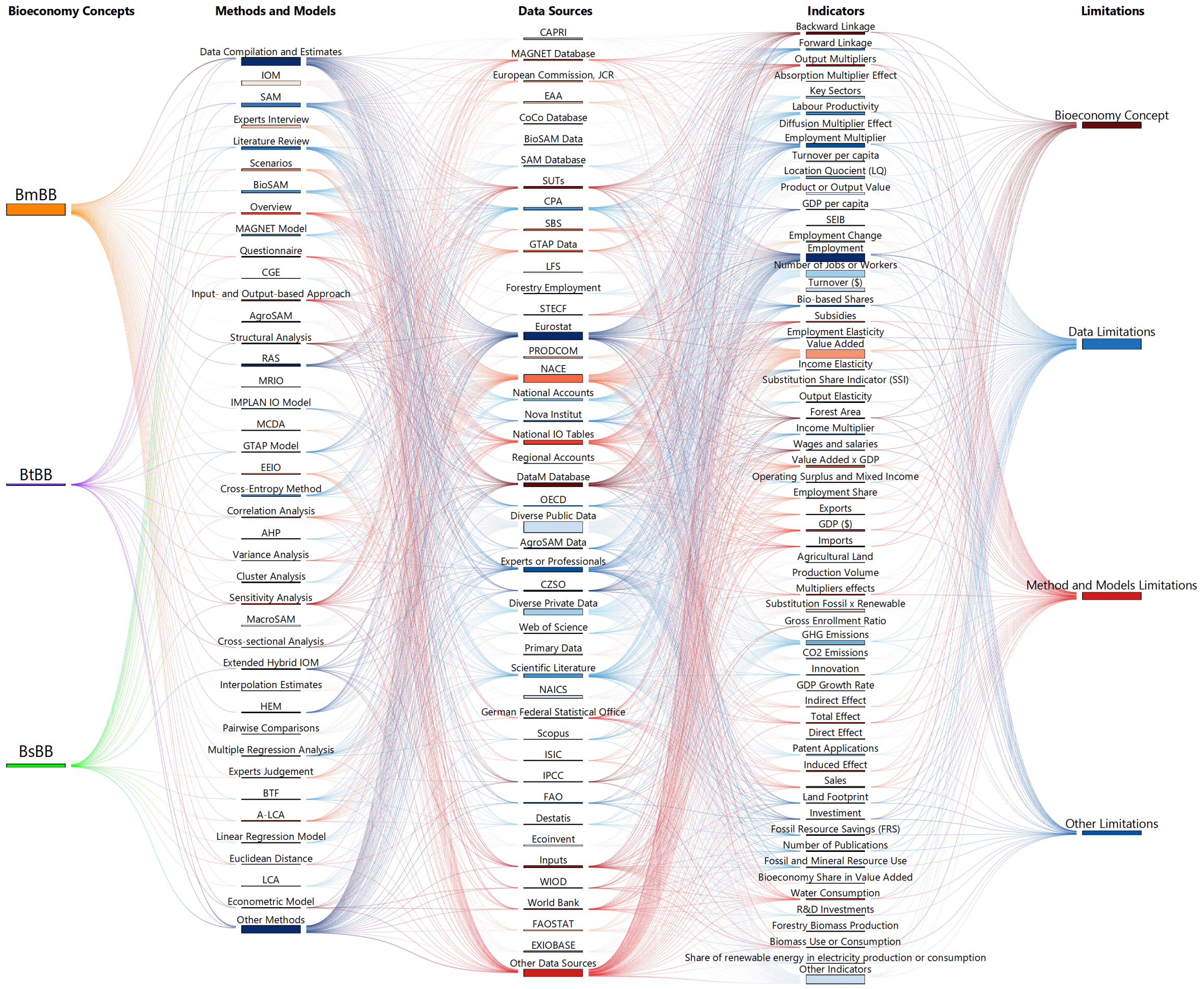

**Figure 5**. An overview of the connections between bioeconomy concepts, methods and models, data sources, indicators and limitations.

Note: The bars' thickness corresponds to the occurrence frequency of the codes. 'Other limitations' include the limitations related to indicators.

Source: elaborated by the authors based on research data.

Figures 6, 7 and 8 explore more clearly the connections between bioeconomy concepts and methods, data and indicators, respectively. A miscellaneous of 72 methods were coded in the documents analyzed. Figure 6 shows those methods with a frequency of two or more occurrences. The methods that occurred only once were grouped under 'Other Methods.' This is the case, for example, with Causality Analysis, Case Studies and Cointegration Tests, among others.

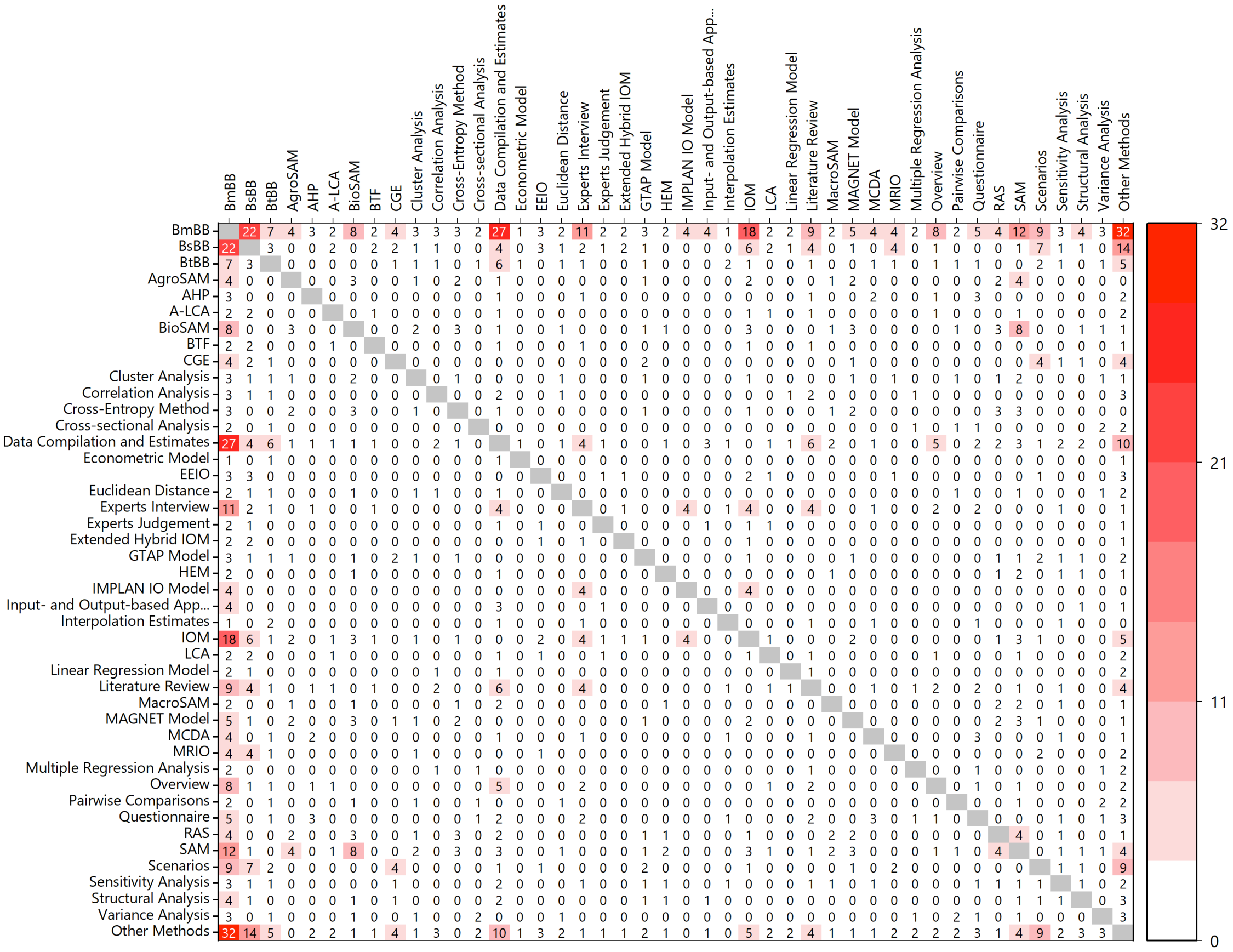

**Figure 6.** The co-occurrence between concepts of bioeconomy and methods.
Source: elaborated by the authors based on research data.

Because the studies focused on measuring the BmBB, most of the methods were connected to this concept of the bioeconomy. Although 'Other Methods' accumulated more frequency, the methods most commonly connected to the BmBB concept were 'Data Compilation and Estimates' (27), 'IOM' (18), 'SAM' (12), 'Experts Interviews' (11), 'Scenarios' and 'Literature Review' (9). However, adding together the methods related to input-output models would be the most frequently used method in studies related to the BmBB.

Most frequently connected to the BsBB concept of bioeconomy are 'Other Methods' (14), 'Scenarios' (7), 'IOM' (6), 'Data Compilations and Estimates' (4), and 'Literature Review' (4). The measurement of the bioeconomy associated with the BtBB concept is connected to the 'Data Compilation and Estimates' (6) and 'Other Methods' (5) methods.

In addition to the results highlighted, it is possible to identify the co-occurrence of methods. In other words, to identify combined methods to measure the bioeconomy. By way of example, we highlight the co-occurrence of the methods 'Data Compilation and Estimates' and 'Other Methods' (10), and 'Scenarios' and 'Others Methods' (9).

Similarly, Figure 7 shows the connections between the concepts and the data or data sources used to measure the bioeconomy. Likewise, the data presented in the figure occurred twice or more. Data that occurred only once has been grouped under 'Other Data Sources.'

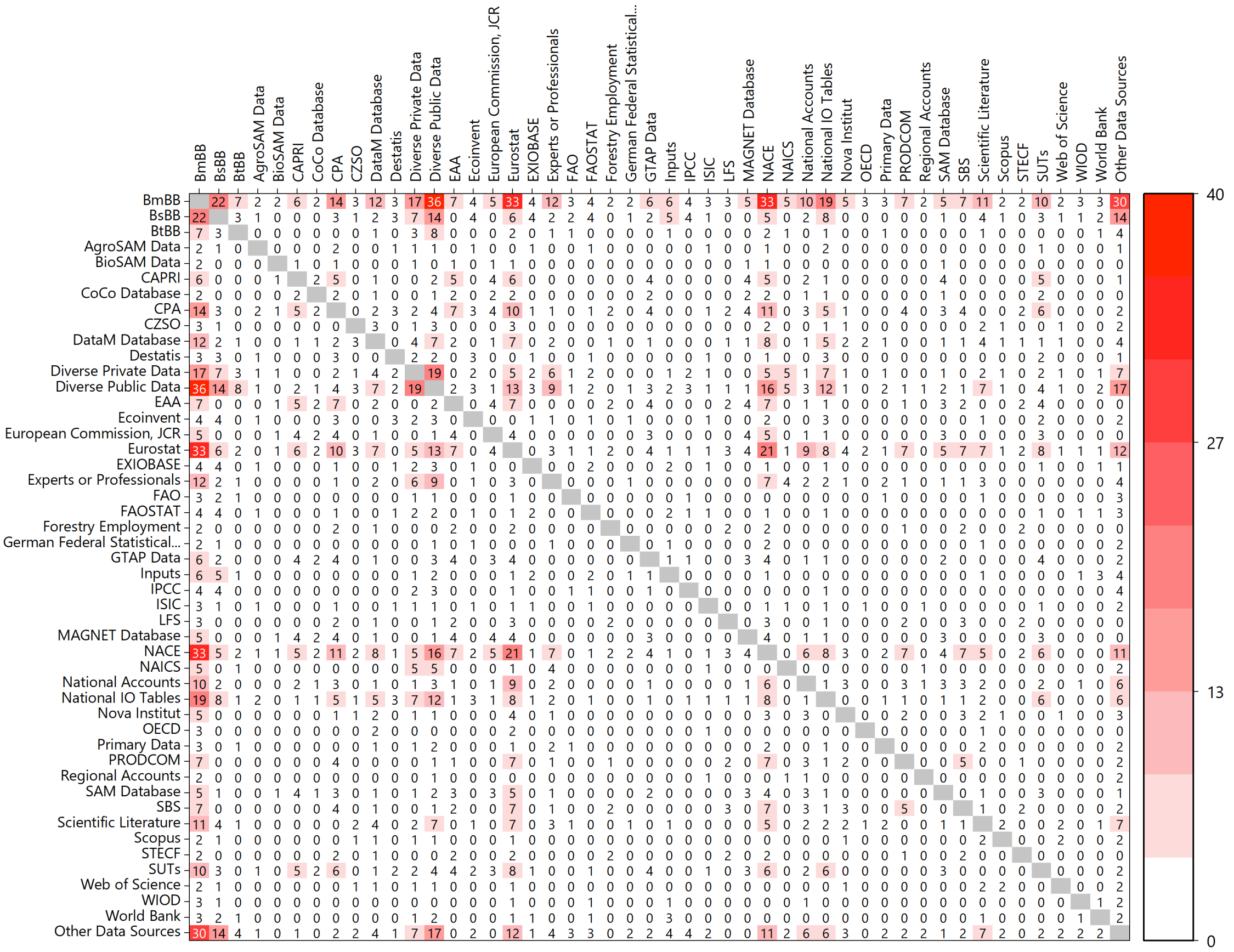

**Figure 7.** The co-occurrence between concepts of bioeconomy and data sources.
Source: elaborated by the authors based on research data.

The main data sources used to measure BmBB were 'Diverse Public Data' (36), 'Eurostat' (33), 'NACE' (33), 'Other Data Sources' (30), and 'National IO Tables' (19). On the other hand, 'Diverse Public Data' (14, 8) and 'Other Data Sources' (14, 4) were the data sources most frequently associated with BsBB and BtBB, respectively. Several co-occurrences of data sources can also be seen in Figure 7, most notably 'Eurostat' and 'NACE' (21), which are a statistical database and an EU activity classification system, respectively.

The connections between the concepts and indicators used to measure the bioeconomy are best shown in Figure 8. By default, only indicators with a frequency of 2 or more are listed, while those with a unit frequency are aggregated under 'Other Indicators.'

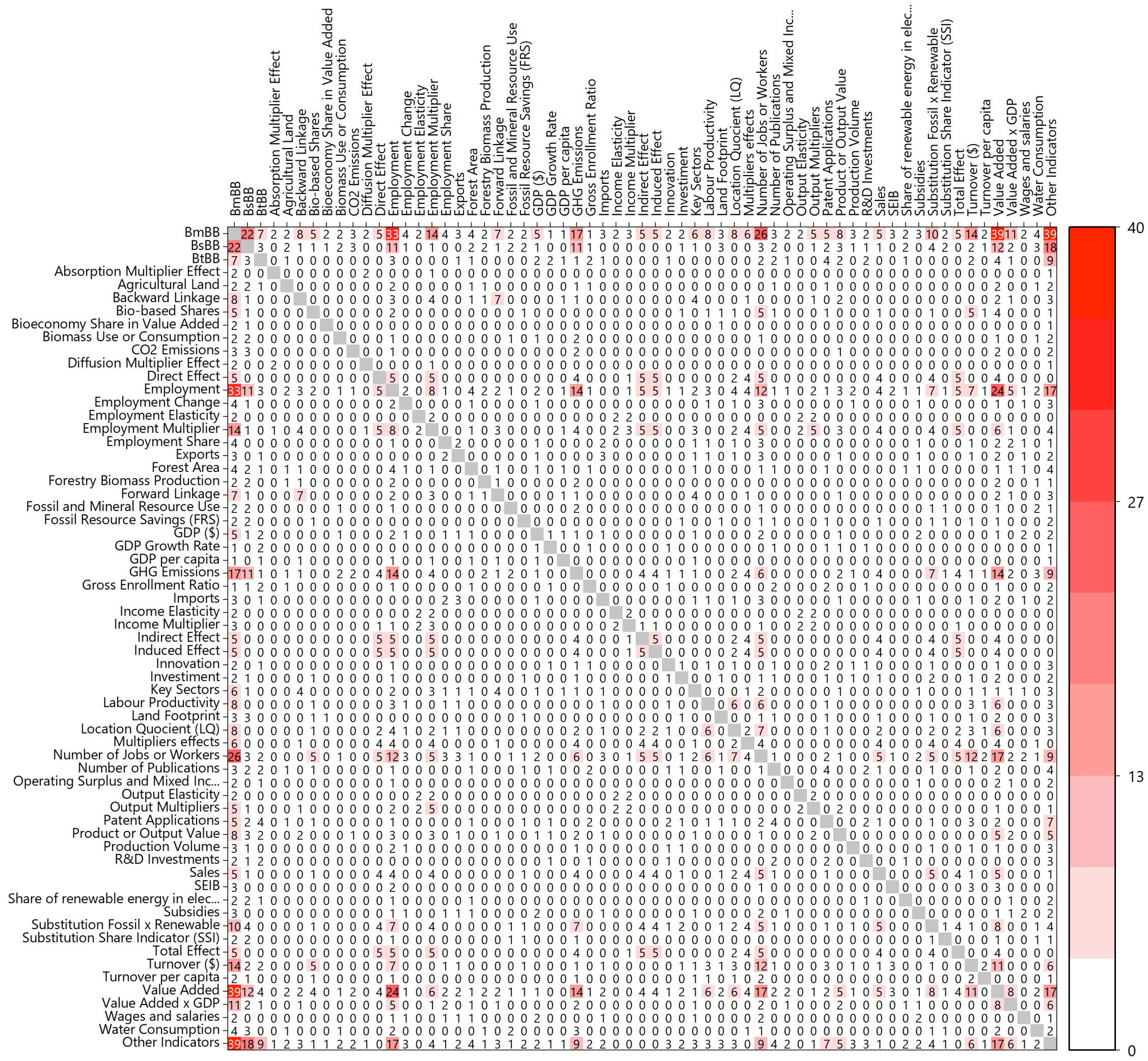

**Figure 8.** The co-occurrence between concepts of bioeconomy and indicators.
Source: elaborated by the authors based on research data.

Around 110 indicators were identified in the documents, 39 of which occurred only once. Among the other 61 indicators, the measurement of 'Value Added' (39), 'Employment' (33), 'Number of Jobs or Workers' (26), ‘GHG Emissions’(17), 'Employment Multiplier' (14) and 'Turnover ($)' (14) were the most frequently connected with the BmBB concept. The indicator labels were chosen to be as faithful as possible to those used in the documents. For this reason, some indicators may measure similar variables, such as 'Employment,' 'Employment Change,' 'Employment Multiplier,' 'Employment Share,' and 'Number of Jobs or Workers,' which could be aggregated in 'Employment.' Therefore, if the indicators associated with the same variable were aggregated, the most frequent indicators

would be related to the socio-economic variables value added, GDP, turnover, and employment.

'GHG Emissions' can be considered the leading indicator associated with the BsBB concept. On the other hand, the BtBB concept appears connected to 'Patent Applications' (4) and 'Other Indicators' (9), such as 'Skilled Labor,' 'RPA' (Revealed Patent Advantage), and 'Public Expenditure in Education.'

Figures 9, 10 and 11 are dedicated to showing in detail the connections between methods and data, data and indicators, and methods and indicators, respectively. As illustrated in Figure 9, there are several co-occurrences between methods and data sources. By way of example, we have highlighted only the most frequent connections. As mentioned above, among the most frequent methods, 'Data Compilation and Estimates' uses numerous data sources, especially 'NACE' and 'Eurostat,' 'Diverse Public Data' and 'Other Data Sources.' Individually, national statistical databases are relevant data sources. Methods related to input-output models (IOM) are linked to 'Diverse Public Data' and 'National IO Tables' as the most used data sources. Also noteworthy are the methods related to the Social Accounting Matrix (SAM), whose most frequent data sources are official EU data, such as 'Eurostat,' 'NACE,' 'SUTs,' 'CPA,' and 'CAPRI,' among others.

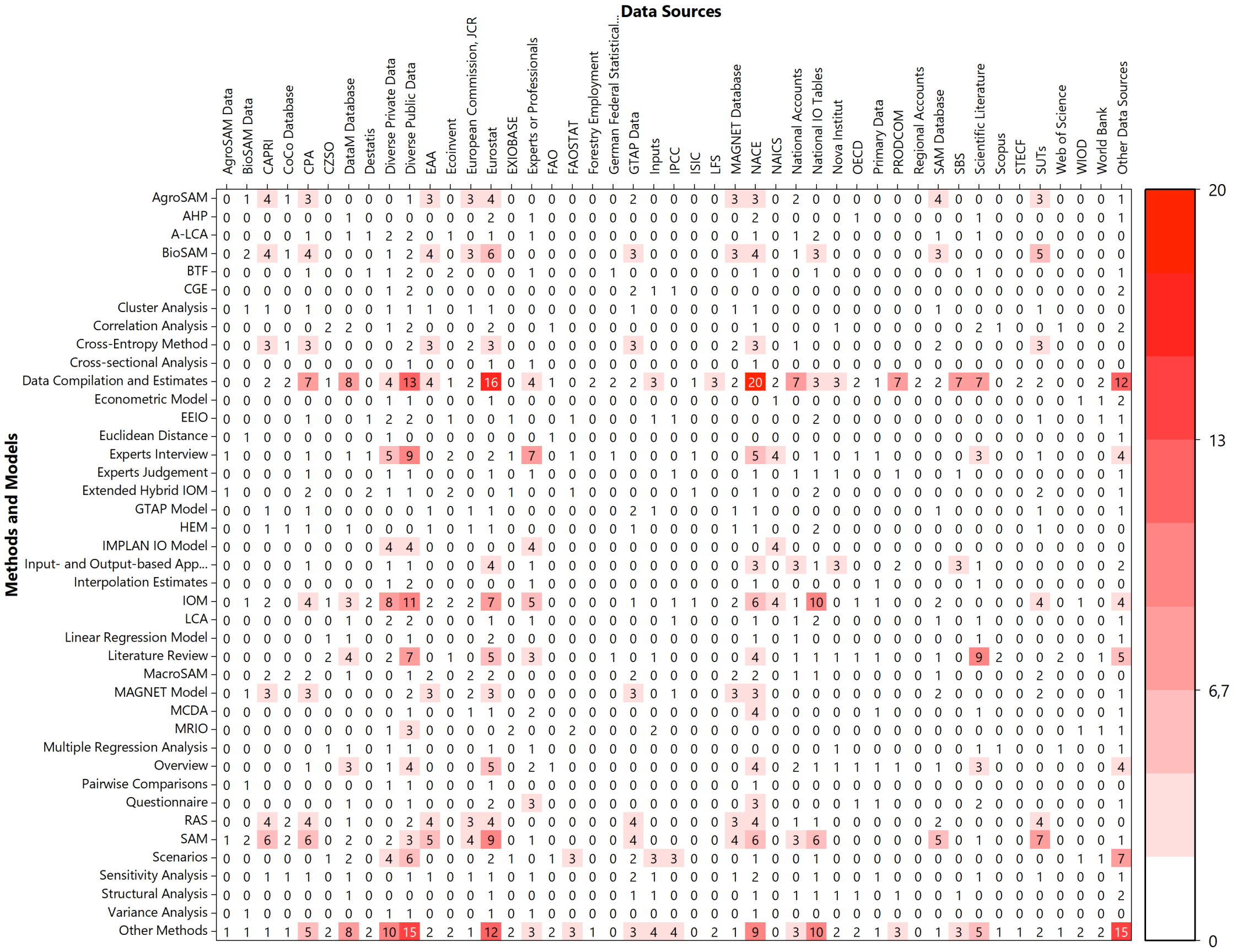

**Figure 9.** The co-occurrence between methods and data sources.
Source: elaborated by the authors based on research data.

The methodological exploration of the data aims to produce results (indicators) that express the bioeconomy share in the national economies. The connections between data sources and indicators can be seen in Figure 10. In general, the data available from the most frequently used data sources, as highlighted in Figures 7 and 9, was used to produce socio-economic indicators using specific methods. Among the most frequent are 'Employment,' 'Value Added,' 'Number of Jobs or Workers,' and 'GHG Emissions.' Many other less frequent indicators and their connection to the specific databases can be identified.

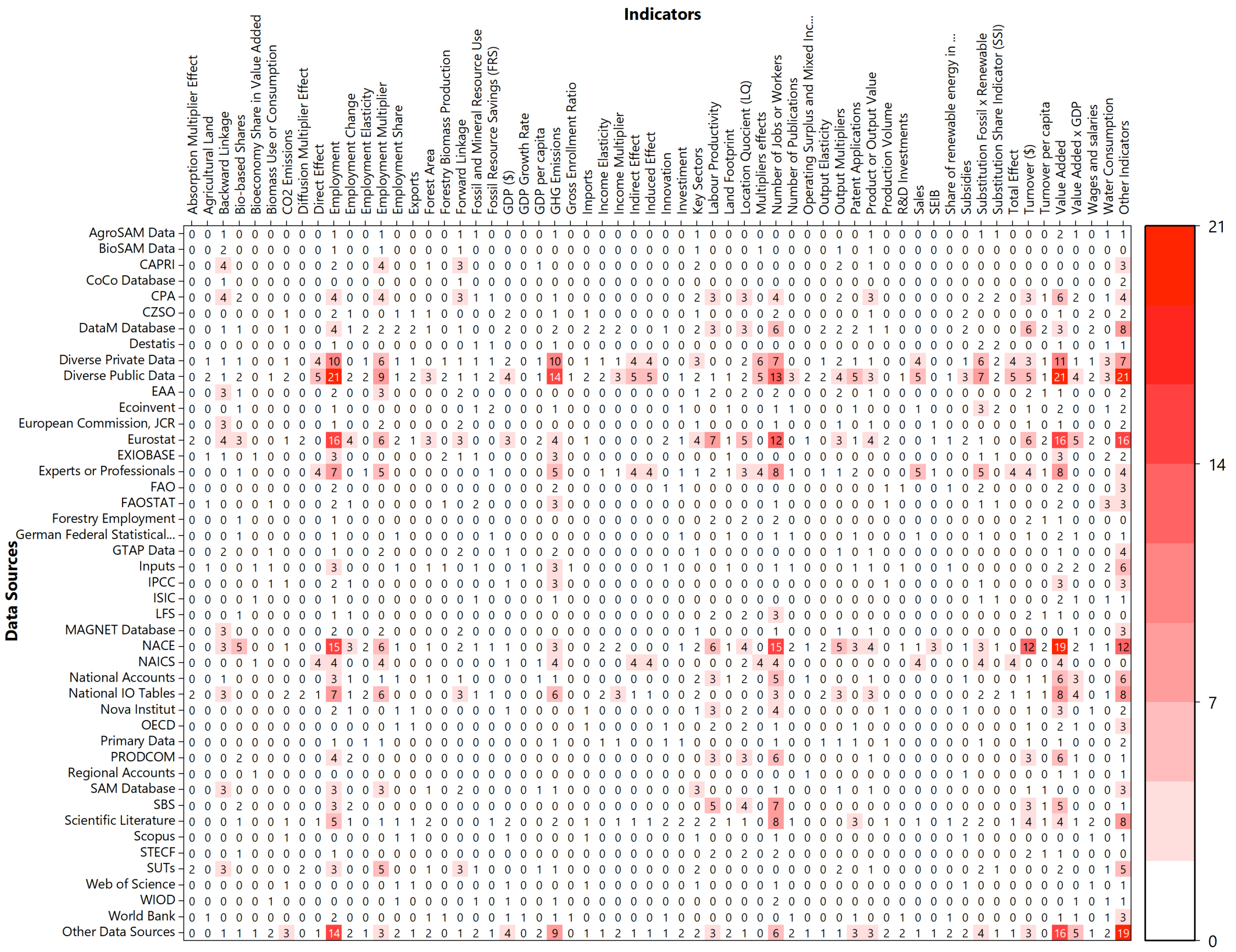

**Figure 10.** The co-occurrence between data sources and indicators.
Source: elaborated by the authors based on research data.

Finally, the connections between the methods used to measure the bioeconomy and the resulting indicators are shown in Figure 11. The most frequently used methods, 'Data Compilation and Estimates,' 'IOM,' and 'SAM,' are connected to the indicators 'Employment,' 'Value Added,' 'Number of Jobs or Workers,' and 'GHG Emissions.'

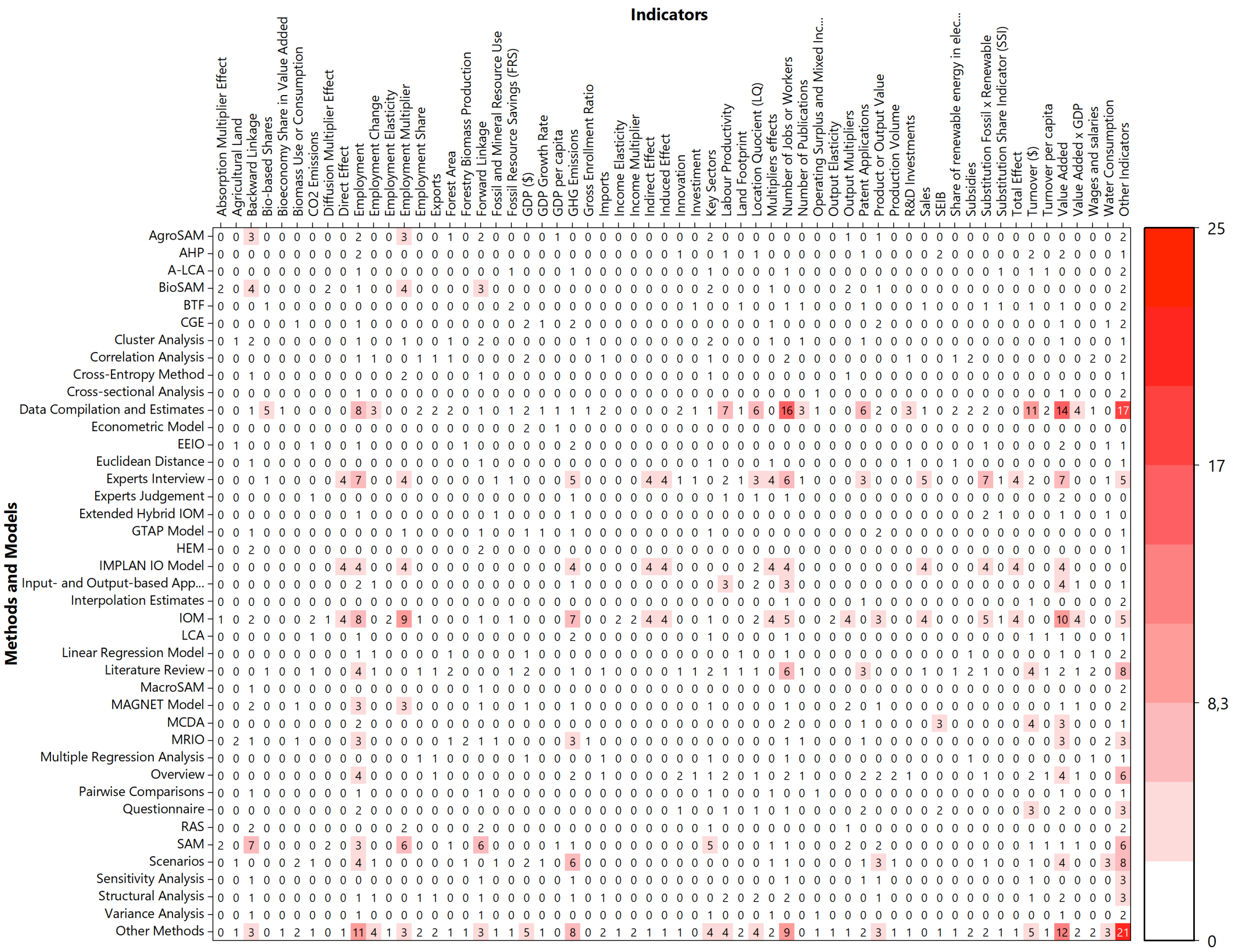

**Figure 11.** The co-occurrence between methods and indicators.
Source: elaborated by the authors based on research data.

When analyzing the content of the documents, an effort was made to identify limitations reported by the authors associated with measuring the bioeconomy. In this process, 71 unique limitations were identified and grouped by those related to the bioeconomy concept, data, method, indicators and others, as shown in Figure 12.

Data limitations were the most frequently reported, accounting for 36.6% of all unique limitations. In this sense, incomplete data, missing or lacking data, aggregated data, outdated data or databases, unavailable data, scarce or limited data, and data quality were the most frequent limitations.

The second position regarding limitations is those associated with methods or models. Twenty-five unique limitations (35.2%) were found in this category. The most frequently reported method limitations were the uncertainty of the estimated values, the linearity assumptions of the models, difficulties in combining models, the subjectivity of some weighting procedures, the absence of supply restrictions in the models, and the assumption of fixed prices.

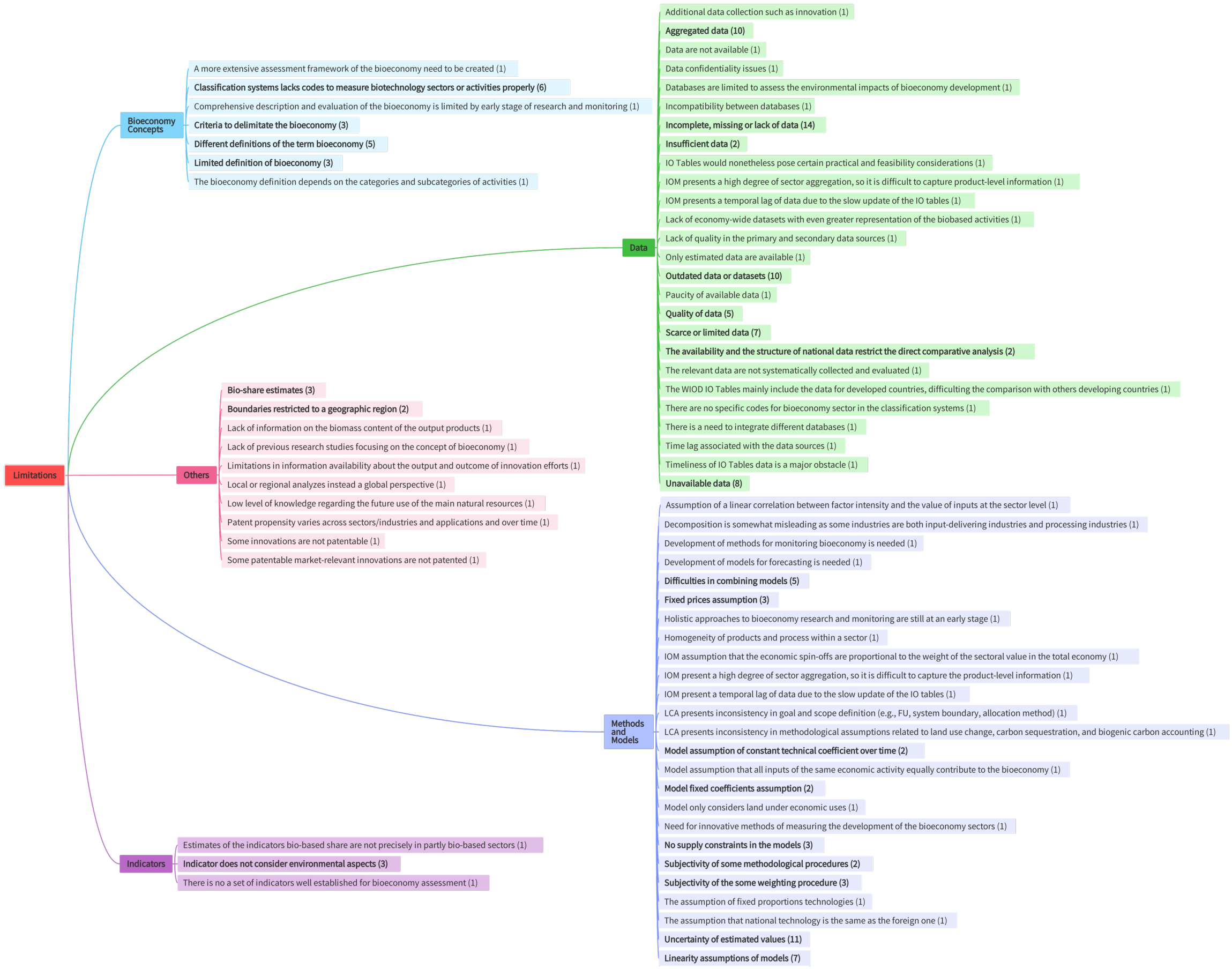

**Figure 12.** The mind map of limitations in measuring the bioeconomy related to the concepts, data, method and models, indicators, and others.
Note: The numbers in parentheses refer to the frequency. The most frequent limitations are in bold.
Source: elaborated by the authors based on research data.

Together, the limitations of methods or models and data were the most relevant and accounted for more than 70% of the unique limitations. Among the remaining limitations, the category of other limitations and bioeconomy concepts account for 23.9% of the unique limitations. In this sense, the main limitations related to the concept of the bioeconomy concern the lack of specific codes in classification systems to measure biotechnology activities or sectors correctly, the existence of different definitions for the term bioeconomy, criteria used to delimit the scope of the bioeconomy to be measured, and the limited definition of the bioeconomy. On the other hand, the other most frequent limitations are associated with the criteria or lack of precision in bio-share estimates, the geographical scope of studies restricted to the borders of a country or region, and limitations related to the measurement of biotechnology (BtBB).

The category of limitations related to indicators was the least relevant, with only three unique limitations. However, these limitations are important to note, as they signal concern about environmental aspects (BsBB), recognize the limitations of the indicator results due to the limitations of the criteria for weighting the participation of biological resources in economic activities and signal the need to develop specific indicators to assess the bioeconomy (BsBB).

## 5. Discussion

### *5.1. Concepts, methods, data, indicators and limitations in measuring the bioeconomy*

The impact of the bioeconomy on national economies has led to various efforts to measure its activities and support public policy planning and decision-making. The findings presented in this study make relevant contributions to the efforts to measure national bioeconomies and point to critical limitations that need to be overcome. This section discusses (i) bioeconomy concepts, (ii) methods or models, (iii) data sources, (iv) indicators and (v) limitations involved in efforts to measure the economic dimension of the bioeconomy.

As elaborated in Section 2, we recognize the existence of at least three conceptual visions of the bioeconomy: BmBB, BtBB, and BsBB, and we consider that the holistic bioeconomy results from the sum of the three concepts. This perspective is shared by Highfill and Chambers (2023). However, only a few studies recognized the three partial concepts, like Alviar et al. (2021), but are limited in measuring holistic bioeconomy. The recurring concept is restricted and primarily based on the BmBB. This may be due to (i) the concentration of studies in EU member countries which adopt a formal concept of the bioeconomy that refers mainly to the bio-based economy; (ii) the possible lack of knowledge or disregard of other concepts due to possible methodological difficulties; or (iii) the objective and scope of the study.

The BsBB concept was considered by the authors of approximately one-quarter of the studies but not explicitly grounded in ecological economics. Essentially, the studies are restricted to combining the measurement of specific indicators of environmental impacts or resource use combined with socioeconomic performance, such as Asada et al. (2020) and Haddad, Britz, and Börner (2019). Although biotechnology has been advocated as an inclusive and sustainable technological platform for the bioeconomy (Lokko et al., 2018), few studies have been dedicated to measuring the economic impacts of BtBB. Carlson (2016) was the only study that dealt exclusively with BtBB, using data compilation to estimate the contribution of biotechnology to the North American economy. Possibly, the lack of structured data that adequately informs the contributions of BtBB is a limitation. Given this, the few studies considering BtBB have mainly explored the turnover of biotech industries or aspects related to patent registration. Besides this, human resources training and R&D investments have been used to evaluate knowledge capital, generally associated with Knowledge-Based Bioeconomy–KBBE (Kircher et al., 2022).

The measurement of the bioeconomy has been based on numerous methods or models, mainly various data compilation and estimation techniques, input-output models (IOMs) and social accounting matrices (SAMs). Macroeconomic studies exploring economic systems' structure and structural changes often resort to IOMs. While IOMs measure the interconnections between economic sectors, SAMs are an expansion of IOMs that add other agents such as government and households (Slovachek, 2023a). Together, IOMs and SAMs make it possible to calculate various dimensions of economic activity, such as intermediate and final demand, technical multipliers, backward and forward linkages, among others.

The studies by Heijman (2016), T. Ronzon et al. (2017), Tévécia Ronzon et al. (2017), Tévécia Ronzon and M'Barek (2018), Tévécia Ronzon et al. (2020), Jurga, Loizou, and Rozakis (2021), Cingiz et al. (2021), Tévécia Ronzon, Iost, and Philippidis (2022a), Tévécia Ronzon, Iost, and Philippidis (2022b) and Lazorcakova et al. (2022), for example, used input-output models (IOMs). Variations on IOMs, such as the Multiregional Input-Output Model (MRIO) and Environmentally-Extended Input-Output Model (EEIO) or Environmentally-Extended Multiregional Input-Output Model (EEMRIO), were used by Asada et al. (2020) to aggregate data and measure environmental aspects associated with the bioeconomy, while Wen et al. (2019) combined Principal Component Analysis (PCA), Autoregressive Distributed Lag (ARDL) and IOM to identify the key sectors and the short- and long-term effects of the bioeconomy on Japan's GDP.

Measuring the bioeconomy using traditional macroeconomic methods can be justified by using the concept of the bioeconomy restricted to the BmBB, where biomass is the main input and primary production sectors are the key sectors. This sectoral structure makes it possible to use data from national input-output tables and measure the participation of the bioeconomy like any other economic activity. In this sense, the measurement of the BmBB differs little from what is usually measured as agribusiness.

Defining the fundamental strategies to be used in measuring the bioeconomy depends on the concept, the method, the geographical scope, and, above all, the availability of data. The unavailability of suitable data has been an obstacle to adequately measuring the bioeconomy. Part of this difficulty is related to the conventional structure of the organization of national accounts, where there is no classification of sectors and activities specific to the bioeconomy and, as stated by Wesseler and von Braun (2017), bioeconomy cuts across sectors and therefore cannot be treated as a traditional sector in economics. Consequently, from the perspective of conventional economic analysis, efforts have been aimed at identifying sectors and activities related to the bioeconomy.

National classification systems for economic sectors and activities have been used alone or in combination to delimit the scope of data representative of the bioeconomy. Thus, coding structures for sectors and activities, such as the Statistical Classification of Economic Activities–NACE (Nowicki et al. 2008a, T. Ronzon et al. 2017, Tévécia Ronzon et al. 2017, Tévécia Ronzon and M'Barek 2018, Jurga, Loizou, and Rozakis 2021, Lazorcakova et al. 2022, Tévécia Ronzon, Iost, and Philippidis 2022b), Classification of the Functions of Government–COFOG, Common International Classification of Ecosystem Services–CICES (Tévécia Ronzon, Iost, and Philippidis 2022a), Statistical Classification of Products by Activity in the European Economic Community–PRODCOM (Asada et al. 2020), Classification of Products–CPA (Lazorcakova et al. 2022), in the EU, and the National Income and Product Accounts–NIPAs, North American Industry Classification System–NAICS, North American Product Classification System–NAPCS (Daystar et al. 2020, Highfill and Chambers 2023), IMPLAN Industry Scheme, in the US are examples of classification system used in delimiting bioeconomy.

Another strategy to better delimitate the bioeconomy is to estimate the relative share of biomass (bio-share) in the economy's sectors, activities or products. Researchers have resorted to sectoral statistical data such as ISTAT Exploring regional transitions to the bioeconomy using a socio-economic indicator: the case of Italy (D'Adamo et al., 2022), PRODCOM (Robert et al., 2020), NACE (Piotrowski, Carus, and Carrez 2018, Piotrowski, Carus, and Carrez 2019, Tévécia Ronzon et al. 2017b, Tévécia Ronzon and M'Barek 2018), or by combining statistical data with the biomass content in products reported in the scientific literature (Ronzon et al., 2022a) or the opinion of sectoral experts collected through interviews. Alternatively, Lazorcakova et al. (2022) estimated minimum, maximum and middle bio-share based on the classification of products provided by the CPA. However, as there is no official data on the subject, determining bio-shares is just as tricky as classifying activities, leaving researchers to use imprecise estimates that compromise the accuracy of the estimated values.

Dozens of indicators have been identified to measure the contributions of the bioeconomy to national economies. The measurement of the value added of the bioeconomy has been recurrent in the literature (Heijman 2016, Tévécia Ronzon et al. 2020, Daystar et al. 2020, Cingiz et al. 2021, Alviar et al. 2021, Tévécia Ronzon, Iost, and Philippidis 2022b), as well as the GDP of the bioeconomy (Wen et al. 2019, Tévécia Ronzon, Iost, and Philippidis 2022a). Other studies have considered the turnover of industries related to the bioeconomy (T. Ronzon et al. 2017, Tévécia Ronzon et al. 2017). The socio-economic contributions of the bioeconomy have also been investigated, mainly through the labor market and jobs (Tévécia Ronzon et al. 2017, Tévécia Ronzon and M'Barek 2018, Tévécia Ronzon et al. 2020, Daystar et al. 2020, Tévécia Ronzon, Iost, and Philippidis 2022a, Tévécia Ronzon, Iost, and Philippidis 2022b). The environmental impacts of the bioeconomy have been assessed in terms of GHG emissions (Alviar et al. 2021, Lazorcakova et al. 2022) or natural resources like land (Haddad, Britz, and Börner 2019). These indicators have been calculated and reported in absolute values, such as $ millions, number of employees, and tons of $CO_2$. Or, in relative terms, such as value added/GDP, turnover/GDP, percentage of jobs/total jobs, and growth rate of these variables over time (Heijman 2016, Tévécia Ronzon et al. 2020, Cingiz et al. 2021, Lazorcakova et al. 2022, Tévécia Ronzon, Iost, and Philippidis 2022a, Tévécia Ronzon, Iost, and Philippidis 2022b). In general, the indicators used to express the contribution of the bioeconomy to national economies are those traditionally used to measure other economic activities, considering the methods predominantly used.

Various limitations have been codified and systematized, with those related to methods or models and data being the most recurrent. Many limitations are inherent to the models, such as the assumptions of constant returns to scale, fixed input structure, industry homogeneity, no supply constraints, and fixed prices, among others, of the IOMs (Slovachek, 2023b). In the same vein, the limitations of the data are closely related to the methods and models used. The limited availability of timely input-output tables is pointed out as a critical limitation of IOM data (van Leeuwen et al., 2005).

### *5.2. Challenges to overcome the limitations to measure the bioeconomy economically*

Jander and Grundmann (2019) state that monitoring bioeconomy transitions and their effects can be considered a Herculean task, as they cannot be easily captured using current economic statistics. The findings of this study reinforce and detail the difficulties and limitations associated with this endeavor. However, it is necessary to move forward, recognizing and confronting the challenges. Of course, many challenges could be listed. Below, we highlight some that are relevant.

The first and perhaps most important question is: which bioeconomy will be measured? In this sense, recognizing the existence of different conceptual approaches and moving towards measuring the holistic bioeconomy is fundamental. It is necessary to consider the complexity surrounding the etymology of the word 'bioeconomy.' Efforts to measure the BmBB are valid and commendable but insufficient. Attempts to measure BsBB use a conceptual perspective restricted to environmental impacts but need to develop an economic approach to these impacts. The few studies considering the BtBB concept have come up against inadequate data.

The first challenge inevitably leads to a second: how do you economically measure something that has no market and, therefore, no economic parameters? This challenge is strongly associated with the BsBB concept. In particular, there is a market for various abiotic natural resources, such as oil, land, and minerals. So, the market parameters (supply, demand, price) are known, and such resources can be incorporated into the usual models without major difficulties. The same is not true of various biotic resources that perform ecosystem services or serve as the basis for biotechnology applications and the production of bioengineered solutions, such as multiple microorganisms living in the soil, air and water.

Carbon dioxide is an example of an abiotic resource facing challenges in the current economic scenario. Due to the growth of global economies based on nonrenewable energy, anthropogenic activity fosters greenhouse gas emissions, including $CO_2$, $CH_4$, and $N_2O$ (Wu et al., 2024). As a result, the World's climate change discussions from the IPCC are crucial. To ensure the sustainable development of countries, a carbon market is necessary (Piris-Cabezas et al., 2023). However, due to the lack of global market regulation, incorporating carbon accounting in bioeconomic measurement is challenging. Additionally, it is difficult to incorporate the value of green biotechnologies related to carbon sequestration in bioeconomic measurement. Although only biotic resources fall within the scope of the bioeconomy, abiotic resources are essential as life substrates. The economic measurement of the holistic bioeconomy can be disregarded if *homo economicus* is assumed truly dead (Gowdy and Iorgulescu Polimeni 2005, Fleming 2017).

The second challenge gives rise to a third one, which lies at the interface between BsBB and BtBB: how do we establish the proportionality of the contribution between biological organisms (bio-share) and knowledge (knowledge-share) in BtBB? Many biotechnological solutions result from applying accumulated human knowledge and physical resources to biological organisms to produce a desired result: food, plant, or medicine. The book 'The Genesis Machine: Our Quest to Rewrite Life in the Age of Synthetic Biology' by Webb and Hessel (2022) and the microbial cell factory approach (Villaverde, 2010) illustrate this relationship well. Would the new route to insulin production be possible only with the knowledge used to map the genetic code? Could the *Escherichia coli* produce insulin voluntarily? Neither. The production of synthetic insulin required the knowledge to map the *Escherichia coli* genome and genetically modify the bacteria to express insulin production through synthetic biology. The economic value and, therefore, the market price of insulin is known, but the bio-share and knowledge-share are unknown. There are many other examples in this direction. The proposition of a system dynamics model for the so-called 'biotechonomy' was proposed by A. Blumberga et al. (2018), but there is much more to be done.

A fourth challenge is the inclusion of other sectors and economic activities beyond the primary ones. Most studies have measured the bioeconomy based on a restricted set of primary sectors, especially agriculture, livestock, fishing and forestry. This section assumes that biomass is the bioeconomy's primary input. Although biomass is relevant, other equally valuable biological inputs are produced and used in many sectors. In this sense, sectors such as the biochemical industry, pharmaceuticals, human and animal health, marine, nanobiotechs, and environmental protection are rarely mentioned. However, they should be considered when measuring the holistic bioeconomy.

Lastly, and the most daunting challenge, is integrating efforts to measure the holistic bioeconomy. The limitations of the methods, data and indicators used in the restricted measurement of the bioeconomy indicate the difficulty in developing methods, producing data and information, and generating indicators capable of measuring the holistic bioeconomy. The challenge increases when it comes to measuring the bioeconomy economically. In this sense, the growing applications of digitalization and computational resources may be the only way to overcome these challenges.

### *5.3. Limitations of the study*

The results reported in the current study were influenced by the scope defined in the purpose and the methodological choices. Consequently, the limitations inherent in the process may have interfered in some way with the outcomes. The main limitations are highlighted below.

Firstly, the study focused on measuring the economic contribution of the bioeconomy. This delimitation left out other measurements related to the bioeconomy, such as circularity and sustainability in its economic, environmental and social dimensions. The literature review by Ferreira et al. (2022) presents methods and indicators related to the sustainability of the bioeconomy, including the economic dimension. However, our results are more detailed and include limitations.

The snowballing strategy for identifying and selecting documents for content analysis may have excluded other documents that, due to variables such as language or journal indexing base, were not cited in the literature consulted. Examples of this are the works by Trigo et al. (2015) for the Argentine bioeconomy and by Silva, Pereira, and Martins (2018), Cicero Zanetti Lima and Pinto (2022) and Cícero Zanetti Lima and Pinto (2022) for the Brazilian bioeconomy. Possibly, other similar studies around the world could have been retrieved, expanding the number of countries measuring their bioeconomy in Figure 4. Therefore, choosing other search strategies could have broadened the spectrum of documents. On the other hand, the studies cited used concepts, methods, data and indicators identified in this study.

No predefined structure was used to code the concepts, methods, data, indicators and limitations. Instead, the possibility of adding codes as they were identified in the documents was kept open. In addition, we tried to adopt a nomenclature that was as faithful as possible to that used by the authors. These choices may have resulted in a long list of codes; some could have been aggregated, resulting in different values. This is the case, for example, with the codes 'IOM,' 'MRIO' and 'EEIO,' which, alternatively, could have been merged into the code 'IOM.'

The subjectivity of the researcher can bias qualitative studies. For this reason, it would be advisable for two or more researchers to code the documents. The QDA Miner software has resources for coding by a team of researchers and statistically comparing the coding results. However, due to restrictions on the number of licenses for the software and the researchers' access to the licensed equipment, coding was only carried out by one senior researcher.

## 6. Conclusions

The current study sought to define the concepts of bioeconomy and to explore the connections between concepts, methods, data and indicators when measuring the bioeconomy economically and the limitations involved in this process. A literature review was used to define the concepts of bioeconomy, while a set of 84 documents were coded regarding the bioeconomy concepts, the methods and data used, the indicators calculated and the limitations involved in measuring the bioeconomy. Based on the findings, we concluded that (i) the current concepts of bioeconomy are limited because they are partly focused on biomass, knowledge or natural resources; (ii) to overcome the conceptual limitations, we proposed the concept of holistic bioeconomy as a comprehensive approach capable of integrating all the partial attributes of the bioeconomy and measuring its real contribution. We argue the measurement of holistic bioeconomy from an economic perspective, although many others awareness can be connected to this concept; (iii) studies in which authors at least recognize the existence of different concepts of the bioeconomy are rare; (iv) measurements of the bioeconomy have been almost restricted to the concept of biomass-based bioeconomy; (v) dozens of methods, data sources and indicators had been used to measure the bioeconomy; (vi) the most frequently used methods, data sources and indicators are those traditionally used for macroeconomic modeling, especially input-output models; (vii) several limitations exist in the connections between bioeconomy concepts, methods, data and indicators; (viii) the results may be biased by the purpose of the study, the methodological design and the concentration of studies carried out by research groups from EU member countries.

The measurement of the bioeconomy in different countries should be encouraged, especially in those where the bioeconomy has great potential. In addition, efforts should be made to overcome existing limitations and qualify bioeconomy measurements, making the results more accurate. However, it is necessary to recognize the existence of distinct conceptual approaches not only in the scientific literature but, fundamentally, in the real bioeconomy. A globally standardized definition of the concepts and methods related to measuring the three types of bioeconomy, or preferably the holistic bioeconomy, will enable comparative studies between countries and the proposal of global public policies aimed at sustainable development. This is the first step towards developing conceptual frameworks that can overcome the limitations listed in this study and the complexity inherent in the process of measuring a holistic bioeconomy. We argue that integrating the digitization efforts underway in various segments of human-nature activities could be one, if not the only, way in this direction.

## References

Allain, S., Ruault, J.-F., Moraine, M., Madelrieux, S., 2022. The 'bioeconomics vs bioeconomy' debate: Beyond criticism, advancing research fronts. Environ. Innov. Soc. Transitions 42, 58–73. https://doi.org/10.1016/j.eist.2021.11.004

Alviar, M., García-Suaza, A., Ramírez-Gómez, L., Villegas-Velásquez, S., 2021. Measuring the Contribution of the Bioeconomy: The Case of Colombia and Antioquia. Sustainability 13, 2353. https://doi.org/10.3390/su13042353

Asada, R., Cardellini, G., Mair-Bauernfeind, C., Wenger, J., Haas, V., Holzer, D., Stern, T., 2020. Effective bioeconomy? a MRIO-based socioeconomic and environmental impact assessment of generic sectoral innovations. Technol. Forecast. Soc. Change 153, 119946. https://doi.org/10.1016/j.techfore.2020.119946

Badampudi, D., Wohlin, C., Petersen, K., 2015. Experiences from using snowballing and database searches in systematic literature studies, in: Proceedings of the 19th International Conference on Evaluation and Assessment in Software Engineering. ACM, New York, NY, USA, pp. 1–10. https://doi.org/10.1145/2745802.2745818

Befort, N., 2020. Going beyond definitions to understand tensions within the bioeconomy: The contribution of sociotechnical regimes to contested fields. Technol. Forecast. Soc. Change 153, 119923. https://doi.org/10.1016/j.techfore.2020.119923

Blumberga, A., Bazbauers, G., Davidsen, P.I., Blumberga, D., Gravelsins, A., Prodanuks, T., 2018. System dynamics model of a biotechonomy. J. Clean. Prod. 172, 4018–4032. https://doi.org/10.1016/j.jclepro.2017.03.132

Blumberga, D., Muizniece, I., Blumberga, A., Baranenko, D., 2016. Biotechonomy Framework for Bioenergy Use. Energy Procedia 95, 76–80. https://doi.org/10.1016/j.egypro.2016.09.025

Bracco, S., Calicioglu, O., Gomez San Juan, M., Flammini, A., 2018. Assessing the Contribution of Bioeconomy to the Total Economy: A Review of National Frameworks. Sustainability 10, 1698. https://doi.org/10.3390/su10061698

BRASIL, 2019. Programa Bioeconomia Brasil (Sociobiodiversidade) [WWW Document]. Catálogo de Políticas Públicas. URL https://catalogo.ipea.gov.br/politica/559/programa-bioeconomia-brasil-sociobiodiversidade (accessed 2.2.24).

Bugge, M., Hansen, T., Klitkou, A., 2016. What Is the Bioeconomy? A Review of the Literature. Sustainability 8, 691. https://doi.org/10.3390/su8070691

Carlson, R., 2016. Estimating the biotech sector's contribution to the US economy. Nat. Biotechnol. 34, 247–255. https://doi.org/10.1038/nbt.3491

Cingiz, K., Gonzalez-Hermoso, H., Heijman, W., Wesseler, J.H.H., 2021. A Cross-Country Measurement of the EU Bioeconomy: An Input–Output Approach. Sustainability 13, 3033. https://doi.org/10.3390/su13063033

D'Adamo, I., Falcone, P.M., Imbert, E., Morone, P., 2022. Exploring regional transitions to the bioeconomy using a socio-economic indicator: the case of Italy. Econ. Polit. 39, 989–1021. https://doi.org/10.1007/s40888-020-00206-4

Dafermos, Y., Nikolaidi, M., Galanis, G., 2017. A stock-flow-fund ecological macroeconomic model. Ecol. Econ. 131, 191–207. https://doi.org/10.1016/j.ecolecon.2016.08.013

Daystar, J., Handfeld, R.B., Pascual-Gonzalez, J., McConnell, E., Golden, J.S., 2020. An Economic Impact Analysis of the U.S. Biobased Products Industry: 2019 Update.

EC, 2018. Review of the 2012 European Bioeconomy Strategy. Brussels. https://doi.org/https://data.europa.eu/doi/10.2777/086770

Efken, J., Banse, M., Rothe, A., Dieter, M., Dirksmeyer, W., Al., E., 2012. Volkswirtschaftliche Bedeutung der biobasierten Wirtschaft in Deutschland (No. Working Paper 07/2012). Braunschweig.

EU, 2013. A bioeconomy strategy for Europe: working with nature for a more sustainable way of living. https://doi.org/https://data.europa.eu/doi/10.2777/17708

EU, 2012. Innovating for sustainable growth – A bioeconomy for Europe. Brussels. https://doi.org/https://data.europa.eu/doi/10.2777/6462

Eversberg, D., Koch, P., Lehmann, R., Saltelli, A., Ramcilovic-Suominen, S., Kovacic, Z., 2023. The more things change, the more they stay the same: promises of bioeconomy and the economy of promises. Sustain. Sci. 18, 557–568. https://doi.org/10.1007/s11625-023-01321-4

Felizardo, K.R., da Silva, A.Y.I., de Souza, É.F., Vijaykumar, N.L., Nakagawa, E.Y., 2018. Evaluating strategies for forward snowballing application to support secondary studies updates, in: Proceedings of the XXXII Brazilian Symposium on Software Engineering. ACM, New York, NY, USA, pp. 184–189. https://doi.org/10.1145/3266237.3266240

Ferreira, V., Fabregat-Aibar, L., Pié, L., Terceño, A., 2022. Research trends and hotspots in bioeconomy impact analysis: a study of economic, social and environmental impacts. Environ. Impact Assess. Rev. 96, 106842. https://doi.org/10.1016/j.eiar.2022.106842

Ferreira, V., Pié, L., Mainar-Causapé, A., Terceño, A., 2023. The bioeconomy in Spain as a new economic paradigm: the role of key sectors with different approaches. Environ. Dev. Sustain. https://doi.org/10.1007/s10668-022-02830-5

Fleming, P., 2017. The Death of Homo Economicus. Pluto Press. https://doi.org/10.2307/j.ctt1v2xw07

Frisvold, G.B., Moss, S.M., Hodgson, A., Maxon, M.E., 2021. Understanding the U.S. Bioeconomy: A New Definition and Landscape. Sustainability 13, 1627. https://doi.org/10.3390/su13041627

Gardossi, L., Philp, J., Fava, F., Winickoff, D., D'Aprile, L., Dell'Anno, B., Marvik, O.J., Lenzi, A., 2023. Bioeconomy national strategies in the G20 and OECD countries: Sharing experiences and comparing existing policies. EFB Bioeconomy J. 3, 100053. https://doi.org/10.1016/j.bioeco.2023.100053

Giampietro, M., 2024. From input–output analysis to the quantification of metabolic patterns: David Pimentel's contribution to the analysis of complex environmental problems. Environ. Dev. Sustain. https://doi.org/10.1007/s10668-023-04400-9
Giampietro, M., 2019. On the Circular Bioeconomy and Decoupling: Implications for Sustainable Growth. Ecol. Econ. 162, 143–156. https://doi.org/10.1016/j.ecolecon.2019.05.001
Giampietro, M., Funtowicz, S.O., 2020. From elite folk science to the policy legend of the circular economy. Environ. Sci. Policy 109, 64–72. https://doi.org/10.1016/j.envsci.2020.04.012
Giuntoli, J., Robert, N., Ronzon, T., Sanchez Lopez, J., Follador, M., Girardi, I., Barredo Cano, J., Borzacchiello, M., Sala, S., M'Barek, R., La Notte, A., Becker, W., Mubareka, S., 2020. Building a monitoring system for the EU bioeconomy. Luxembourg. https://doi.org/10.2760/717782
Golden, J.S., Handfield, R.B., Daystar, J., McConnell, T.E., 2015. An Economic Impact Analysis of the US Biobased Products Industry: A Report to the Congress of the United States of America. Ind. Biotechnol. 11, 201–209. https://doi.org/10.1089/ind.2015.29002.jsg
Gowdy, J.M., Iorgulescu Polimeni, R., 2005. The death of homo economicus : is there life after welfare economics? Int. J. Soc. Econ. 32, 924–938. https://doi.org/10.1108/03068290510623771
Greenhalgh, T., Peacock, R., 2005. Effectiveness and efficiency of search methods in systematic reviews of complex evidence: audit of primary sources. BMJ 331, 1064–1065. https://doi.org/10.1136/bmj.38636.593461.68
Haddad, S., Britz, W., Börner, J., 2019. Economic Impacts and Land Use Change from Increasing Demand for Forest Products in the European Bioeconomy: A General Equilibrium Based Sensitivity Analysis. Forests 10, 52. https://doi.org/10.3390/f10010052
Heijman, W., 2016. How big is the bio-business? Notes on measuring the size of the Dutch bio-economy. NJAS Wageningen J. Life Sci. 77, 5–8. https://doi.org/10.1016/j.njas.2016.03.004
Highfill, T., Chambers, M., 2023. Developing a National Measure of the Economic Contributions of the Bioeconomy. Suitland, MD.
Huang, A., 2008. Similarity measures for text document clustering, in: 6th New Zealand Computer Science Research Student Conference, NZCSRSC 2008. Christchurch, New Zealand, pp. 49–56.
IEA, 2022. Global Biorefinery Status Report 2022. IEA Bioenergy, Paris, France.
Jalali, S., Wohlin, C., 2012. Systematic literature studies, in: Proceedings of the ACM-IEEE International Symposium on Empirical Software Engineering and Measurement. ACM, New York, NY, USA, pp. 29–38. https://doi.org/10.1145/2372251.2372257
Jander, W., Grundmann, P., 2019. Monitoring the transition towards a bioeconomy: A general framework and a specific indicator. J. Clean. Prod. 236, 117564. https://doi.org/10.1016/j.jclepro.2019.07.039
Jander, W., Wydra, S., Wackerbauer, J., Grundmann, P., Piotrowski, S., 2020. Monitoring Bioeconomy Transitions with Economic–Environmental and Innovation Indicators: Addressing Data Gaps in the Short Term. Sustainability 12, 4683. https://doi.org/10.3390/su12114683
Jurga, P., Loizou, E., Rozakis, S., 2021. Comparing Bioeconomy Potential at National vs. Regional Level Employing Input-Output Modeling. Energies 14, 1714. https://doi.org/10.3390/en14061714
Kardung, M., Costenoble, O., Dammer, L., Delahaye, R., Lovrić, M., Leeuwen, M. van, M'Barek, R., Meijl, H. van, Piotrowski, S., Ronzon, T., Verhoog, D., Verkerk, H., Vrachioli, M., Wesseler, J., Zhu, B.X., 2019. Framework for measuring the size and development of the bioeconomy.
Karvonen, J., Halder, P., Kangas, J., Leskinen, P., 2017. Indicators and tools for assessing sustainability impacts of the forest bioeconomy. For. Ecosyst. 4, 2. https://doi.org/10.1186/s40663-017-0089-8
Kircher, M., Maurer, K.-H., Herzberg, D., 2022. KBBE: The knowledge-based bioeconomy: Concept, status and future prospects. EFB Bioeconomy J. 2, 100034. https://doi.org/10.1016/j.bioeco.2022.100034

Kuosmanen, T., Kuosmanen, N., El-Meligli, A., Ronzon, T., Gurria, P., Iost, S., M'Barek, R., 2020a. How Big is the Bioeconomy? Reflections from an economic perspective. Luxembourg. https://doi.org/10.2760/144526
Kuosmanen, T., Kuosmanen, N., El Meligi, A., Ronzon, T., Gurria, P., Iost, S., M'barek, R., 2020b. How big is the bioeconomy? Reflections from an economic perspective. Publications Office of the European Union, Luxembourg. https://doi.org/10.2760/144526
Lahitani, A.R., Permanasari, A.E., Setiawan, N.A., 2016. Cosine similarity to determine similarity measure: Study case in online essay assessment, in: 2016 4th International Conference on Cyber and IT Service Management. IEEE, pp. 1–6. https://doi.org/10.1109/CITSM.2016.7577578
Lazorcakova, E., Dries, L., Peerlings, J., Pokrivcak, J., 2022. Potential of the bioeconomy in Visegrad countries: An input-output approach. Biomass and Bioenergy 158, 106366. https://doi.org/10.1016/j.biombioe.2022.106366
Lier, M., Aarne, M., Kärkkäinen, L., Korhonen, K.T., Yli-Viikari, A., Packalen, T., 2018. Synthesis on bioeconomy monitoring systems in the EU Member States - indicators for monitoring the progress of bioeconomy. Helsinki, Finland.
Lima, Cicero Zanetti, Pinto, T.P., 2022. PIB da Bioeconomia: métodos e relações de oferta, in: 50º Encontro Nacional de Economia. Associação Nacional dos Centros de Pós-Graduação em Economia, ANPEC, Fortaleza, p. 16.
Lima, Cícero Zanetti, Pinto, T.P., 2022. PIB da Bioeconomia. Fundação Getúlio Vargas - FGV-EESP, São Paulo, SP.
Lokko, Y., Heijde, M., Schebesta, K., Scholtès, P., Van Montagu, M., Giacca, M., 2018. Biotechnology and the bioeconomy—Towards inclusive and sustainable industrial development. N. Biotechnol. 40, 5–10. https://doi.org/10.1016/j.nbt.2017.06.005
NASEM, 2020. Safeguarding the Bioeconomy. National Academies Press, Washington, D.C. https://doi.org/10.17226/25525
Nowicki, P., Banse, M., Bolck, C., Bos, H., Scott, E., 2008. Biobased economy: State-of-the-art assessment. Wageningen University Publishing, The Hague.
Odum, H.T., 1996. Environmental Accounting: Emergy and Environmental Decision Making. John Wiley & Sons, Inc., New York.
OECD, 2023. Key biotechnology indicators [WWW Document]. Innov. > Sci. Technol. Innov. policy. URL https://www.oecd.org/innovation/inno/keybiotechnologyindicators.htm (accessed 2.14.24).
OECD, 2018. Meeting Policy Challenges for a Sustainable Bioeconomy. OECD. https://doi.org/10.1787/9789264292345-en
OECD, 2009. The Bioeconomy to 2030. Designing a Policy Agenda. OECD, Paris, France. https://doi.org/10.1787/9789264056886-en
Patermann, C., Aguilar, A., 2021. A bioeconomy for the next decade. EFB Bioeconomy J. 1, 100005. https://doi.org/https://doi.org/10.1016/j.bioeco.2021.100005
Piotrowski, S., Carus, M., Carrez, D., 2019. European Bioeconomy in Figures 2008 – 2016. Hürth.
Piotrowski, S., Carus, M., Carrez, D., 2018. European Bioeconomy in Figures 2008 – 2015. Hürth.
Piris-Cabezas, P., Lubowski, R.N., Leslie, G., 2023. Estimating the potential of international carbon markets to increase global climate ambition. World Dev. 167, 106257. https://doi.org/10.1016/j.worlddev.2023.106257
Priefer, C., Jörissen, J., Frör, O., 2017. Pathways to Shape the Bioeconomy. Resources 6, 10. https://doi.org/10.3390/resources6010010
Robert, N., Jonsson, R., Chudy, R., Camia, A., 2020. The EU Bioeconomy: Supporting an Employment Shift Downstream in the Wood-Based Value Chains? Sustainability 12, 758. https://doi.org/10.3390/su12030758
Ronzon, T., Iost, S., Philippidis, G., 2022a. An output-based measurement of EU bioeconomy services: Marrying statistics with policy insight. Struct. Chang. Econ. Dyn. 60, 290–301. https://doi.org/10.1016/j.strueco.2021.10.005

Ronzon, T., Iost, S., Philippidis, G., 2022b. Has the European Union entered a bioeconomy transition? Combining an output-based approach with a shift-share analysis. Environ. Dev. Sustain. 24, 8195–8217. https://doi.org/10.1007/s10668-021-01780-8

Ronzon, T., Lusser, M., (ed.), M.K., Landa, L., (ed.), J.S.L., M'Barek, R., (ed.), G.H., (ed.), A.B., (ed.), A.C., Giuntoli, J., Cristobal, J., Parisi, C., Ferrari, E., Marelli, L., Matos, C.T. de, Barbero, M.G., Cerez, E.R., 2017. Bioeconomy Report 2016. European Commission, Brussel. https://doi.org/10.2760/20166

Ronzon, T., M'Barek, R., 2018. Socioeconomic Indicators to Monitor the EU's Bioeconomy in Transition. Sustainability 10, 1745. https://doi.org/10.3390/su10061745

Ronzon, Tévécia, Piotrowski, S., M'Barek, R., Carus, M., 2017a. A systematic approach to understanding and quantifying the EU's bioeconomy. Bio-Based Appl. Econ. 6, 1–17. https://doi.org/https://doi.org/10.13128/BAE-20567

Ronzon, Tévécia, Piotrowski, S., M'Barek, R., Carus, M., 2017b. A systematic approach to understanding and quantifying the EU's bioeconomy. Bio-Based Appl. Econ. 6, 1–17. https://doi.org/10.13128/BAE-20567

Ronzon, T., Piotrowski, S., Tamosiunas, S., Dammer, L., Carus, M., M'barek, R., 2020. Developments of Economic Growth and Employment in Bioeconomy Sectors across the EU. Sustainability 12, 4507. https://doi.org/10.3390/su12114507

Ronzon, T., Santini, F., M'Barek, R., 2015. The Bioeconomy in the European Union in numbers. Facts and figures on biomass, turnover and employment. Seville, Spain.

Rosegrant, M.W., Ringler, C., Zhu, T., Tokgoz, S., Bhandary, P., 2013. Water and food in the bioeconomy: challenges and opportunities for development. Agric. Econ. 44, 139–150. https://doi.org/10.1111/agec.12058

Silva, M.F. de O. e, Pereira, F. dos S., Martins, J.V.B., 2018. A Bioeconomia Brasileira em números. BNDES Setorial 47, 277–332.

Slovachek, A., 2023a. Input-Output & Social Accounting Matrix Structure [WWW Document]. IMPLAN - Support > Ref. > A Snapshot Econ. Input-Output SAM Model. URL https://support.implan.com/hc/en-us/articles/18943702175003-Input-Output-Social-Accounting-Matrix-Structure#:~:text=The interconnectedness of industries is,such as Households and Government. (accessed 1.23.24).

Slovachek, A., 2023b. Input-Output Model Assumptions [WWW Document]. Assumptions I-O. URL https://support.implan.com/hc/en-us/articles/18944187743643-Assumptions-of-I-O (accessed 1.23.24).

Trigo, E., Regúnaga, M., Costa, R., Wierny, M., Coremberg, A., 2015. The Argentinean bioeconomy : scope, present state and opportunities for its sustainable development. Bolsa de Cereales de Buenos Aires, Buenos Aires.

US, 2012. National Bioeconomy Blueprint. Washington, DC, USA.

van Leeuwen, E.S., Nijkamp, P., Rietveld, P., 2005. Regional Input–Output Analysis, in: Encyclopedia of Social Measurement. Elsevier, pp. 317–323. https://doi.org/10.1016/B0-12-369398-5/00349-2

Vandermeulen, V., Prins, W., Nolte, S., Van Huylenbroeck, G., 2011. How to measure the size of a bio-based economy: Evidence from Flanders. Biomass and Bioenergy 35, 4368–4375. https://doi.org/10.1016/j.biombioe.2011.08.007

Villaverde, A., 2010. Nanotechnology, bionanotechnology and microbial cell factories. Microb. Cell Fact. 9, 53. https://doi.org/10.1186/1475-2859-9-53

Vivien, F.-D., Nieddu, M., Befort, N., Debref, R., Giampietro, M., 2019. The Hijacking of the Bioeconomy. Ecol. Econ. 159, 189–197. https://doi.org/10.1016/j.ecolecon.2019.01.027

Webb, A., Hessel, A., 2022. The Genesis Machine: Our Quest to Rewrite Life in the Age of Synthetic Biology, 1st ed. PublicAffairs, New York, NY.

Wei, X., Liu, Q., Pu, A., Wang, S., Chen, F., Zhang, L., Zhang, Y., Dong, Z., Wan, X., 2022a. Knowledge Mapping of bioeconomy: A bibliometric analysis. J. Clean. Prod. 373, 133824. https://doi.org/10.1016/j.jclepro.2022.133824

Wei, X., Luo, J., Pu, A., Liu, Q., Zhang, L., Wu, S., Long, Y., Leng, Y., Dong, Z., Wan, X., 2022b. From Biotechnology to Bioeconomy: A Review of Development Dynamics and Pathways. Sustainability 14, 10413. https://doi.org/10.3390/su141610413

Wen, X., Quacoe, Daniel, Quacoe, Dinah, Appiah, K., Ada Danso, B., 2019. Analysis on Bioeconomy's Contribution to GDP: Evidence from Japan. Sustainability 11, 712. https://doi.org/10.3390/su11030712

Wesseler, J., von Braun, J., 2017. Measuring the Bioeconomy: Economics and Policies. Annu. Rev. Resour. Econ. 9, 275–298. https://doi.org/10.1146/annurev-resource-100516-053701

Wohlin, C., 2016. Second-generation systematic literature studies using snowballing, in: Proceedings of the 20th International Conference on Evaluation and Assessment in Software Engineering. ACM, New York, NY, USA, pp. 1–6. https://doi.org/10.1145/2915970.2916006

Wohlin, C., 2014. Guidelines for snowballing in systematic literature studies and a replication in software engineering, in: Proceedings of the 18th International Conference on Evaluation and Assessment in Software Engineering. ACM, New York, NY, USA, pp. 1–10. https://doi.org/10.1145/2601248.2601268

Wohlin, C., Kalinowski, M., Romero Felizardo, K., Mendes, E., 2022. Successful combination of database search and snowballing for identification of primary studies in systematic literature studies. Inf. Softw. Technol. 147, 106908. https://doi.org/10.1016/j.infsof.2022.106908

Wu, Y., Liu, X., Tang, C., 2024. Carbon Market and corporate financing behavior-From the perspective of constraints and demand. Econ. Anal. Policy 81, 873–889. https://doi.org/10.1016/j.eap.2024.01.006

Zeug, W., Kluson, F.R., Mittelstädt, N., Bezama, A., Thrän, D., 2021. Results from a stakeholder survey on bioeconomy monitoring and perceptions on bioeconomy in Germany (No. UFZ Discussion Paper, No. 8/2021). Leipzig.

Zhang, X., Zhao, C., Shao, M.-W., Chen, Y.-L., Liu, P., Chen, G.-Q., 2022. The roadmap of bioeconomy in China. Eng. Biol. 6, 71–81. https://doi.org/10.1049/enb2.12026

## **Appendix A.** List of documents included in the analysis.

**Order and Reference Information**

1. Alviar, Mauricio, Andrés García-Suaza, Laura Ramírez-Gómez, and Simón Villegas-Velásquez. 2021. "Measuring the Contribution of the Bioeconomy: The Case of Colombia and Antioquia" Sustainability 13, no. 4: 2353. https://doi.org/10.3390/su13042353
2. Asada, R. et al. 2020. Effective bioeconomy? A MRIO-based socioeconomic and environmental impact assessment of generic sectoral innovations, Technol. Forecast. Soc. Change, 153, 119946, https://doi.org/10.1016/j.techfore.2020.119946.
3. Biber-Freudenberger, Lisa, Amit Kumar Basukala, Martin Bruckner, and Jan Börner. 2018. "Sustainability Performance of National Bio-Economies" Sustainability 10, no. 8: 2705. https://doi.org/10.3390/su10082705
4. Bringezu, S., Distelkamp, M., Lutz, C. et al. Environmental and socioeconomic footprints of the German bioeconomy. Nat Sustain 4, 775–783 (2021). https://doi.org/10.1038/s41893-021-00725-3
5. Budzinski, M et al. 2017. Monitoring the progress towards bioeconomy using multi-regional input-output analysis: The example of wood use in Germany. Journal of Cleaner Production, v. 161, pp. 1-11. https://doi.org/10.1016/j.jclepro.2017.05.090
6. Capasso, Marco, and Antje Klitkou. 2020. "Socioeconomic Indicators to Monitor Norway's Bioeconomy in Transition" Sustainability 12, no. 8: 3173. https://doi.org/10.3390/su12083173
7. Cardenete, M. A., Boulanger, P., Del Carmen Delgado, M., Ferrari, E., & M'Barek, R. (2014). AGRI-FOOD AND BIO-BASED ANALYSIS IN THE SPANISH ECONOMY USING A KEY SECTOR APPROACH. Review of Urban & Regional Development Studies, 26(2), 112–134. doi:10.1111/rurd.12022
8. Carlson R. 2016. Estimating the biotech sector's contribution to the US economy. Nat. Biotechnol. 34(3):247–55
9. Cingiz, Kutay, Hugo Gonzalez-Hermoso, Wim Heijman, and Justus H. H. Wesseler. 2021. "A Cross-Country Measurement of the EU Bioeconomy: An Input–Output Approach" Sustainability 13, no. 6: 3033. https://doi.org/10.3390/su13063033
10. D'Adamo, I. et al. 2020. A New Socio-economic Indicator to Measure the Performance of Bioeconomy Sectors in Europe. Ecological Economics, v. 176, 106724. https://doi.org/10.1016/j.ecolecon.2020.106724
11. D'Adamo, I., Falcone, P.M., Imbert, E. et al. Exploring regional transitions to the bioeconomy using a socio-economic indicator: the case of Italy. Econ Polit 39, 989–1021 (2022). https://doi.org/10.1007/s40888-020-00206-4
12. Daystar, J. Handfield, R., Golden, J. S., McConnell, E., Pascual-Gonzalez, J. 2021. An Economic Impact Analysis of the US Biobased Products Industry. Industrial Biotechnology, v. 17, n. 5. https://doi.org/10.1089/ind.2021.29263.jda
13. Daystar, J., Handfeld, R.B., Golden, J.S., and McConnell, T.E. "An Economic Impact Analysis of the U.S. Biobased Products Industry: 2018 Update." A Joint Publication of the A Joint Publication of the Supply Chain Resource Cooperative at North Carolina State University and the College of Engineering and Technology at East Carolina University, 2019. https://www.biopreferred.gov/BPResources/fles/BiobasedProductsEconomicAnalysis2018.pdf.
14. Dolge, K., et al. 2023. A Comparative Analysis of Bioeconomy Development in European Union Countries. Environmental Management 71, 215–233. https://doi.org/10.1007/s00267-022-01751-3
15. Efken J, Banse M, Rothe A, Dieter M, Dirksmeyer W, et al. 2012. Volkswirtschaftliche Bedeutung der biobasierten Wirtschaft in Deutschland. Work. Pap., 07/2012, Johann Heinrich v. Th¨ unen-Inst., Braunschweig, Ger. (Deutsch) – Included in the start-set but not analyzed because German language.
16. Efken, J., Walter Dirksmeyer, Peter Kreins & Marius Knecht (2016) Measuring the importance of the bioeconomy in Germany: Concept and illustration, NJAS: Wageningen Journal of Life Sciences, 77:1, 9-17, DOI: 10.1016/j.njas.2016.03.008
17. Ferreira, V. Pi, L., Terceño, A. 2021. Economic impact of the bioeconomy in Spain: Multiplier effects with a bio social accounting matrix. Journal of Cleaner Production, v. 298, 126752. https://doi.org/10.1016/j.jclepro.2021.126752

18. Ferreira, V., Pié, L., Mainar-Causapé, A. et al. The bioeconomy in Spain as a new economic paradigm: the role of key sectors with different approaches. Environ Dev Sustain (2023). https://doi.org/10.1007/s10668-022-02830-5
19. Ferreira, Valeria, Laia Pié, and Antonio Terceño. 2020. "The Role of the Foreign Sector in the Spanish Bioeconomy: Two Approaches Based on SAM Linear Models" International Journal of Environmental Research and Public Health 17, no. 24: 9381. https://doi.org/10.3390/ijerph17249381
20. Fuentes-Saguar, Patricia D., Alfredo J. Mainar-Causapé, and Emanuele Ferrari. 2017. "The Role of Bioeconomy Sectors and Natural Resources in EU Economies: A Social Accounting Matrix-Based Analysis Approach" Sustainability 9, no. 12: 2383. https://doi.org/10.3390/su9122383
21. Gatto, Fabiana, Sara Daniotti, and Ilaria Re. 2021. "Driving Green Investments by Measuring Innovation Impacts. Multi-Criteria Decision Analysis for Regional Bioeconomy Growth" Sustainability 13, no. 21: 11709. https://doi.org/10.3390/su132111709
22. Golden JS,Handfield RB, Daystar J, McConnell TE. 2015. An economic impact analysis of the U.S. biobased products industry: a report to the Congress of the United States of America. Ind. Biotech. 11(4):201–9
23. Golden, J.S., Handfeld, R.B., Daystar, J., Morrison, B., and McConnell, T.E. "An Economic Impact Analysis of the U.S. Biobased Products Industry: 2016 Update." A Joint Publication of the Duke Center for Sustainability & Commerce and the Supply Chain Resource Cooperative at North Carolina State University, 2016. https://www.biopreferred.gov/BPResources/fles/BiobasedProductsEconomicAnalysis2016.pdf.
24. Grealis, Eoin & O'Donoghue, Cathal, 2015. The Economic Impact of the Irish Bio-Economy: Development and Uses, Research Reports 210704, National University of Ireland, Galway, Socio-Economic Marine Research Unit. DOI: 10.22004/ag.econ.210704
25. Haddad, Salwa, Wolfgang Britz, and Jan Börner. 2019. "Economic Impacts and Land Use Change from Increasing Demand for Forest Products in the European Bioeconomy: A General Equilibrium Based Sensitivity Analysis" Forests 10, no. 1: 52. https://doi.org/10.3390/f10010052
26. Heijman W. 2016. How big is the bio-business? Notes on measuring the size of the Dutch bio-economy. NJAS 77:5–8
27. Iost, S.; Labonte, N.; Banse, M.; Geng, N.; Jochem, D.; Schweinle, J.;Weber, S.;Weimar, H. 2019. German Bioeconomy: Economic Importance and Concept of Measurement. Ger. J. Agric. Econ, 68, 275–288.
28. Jander, W. 2021. An extended hybrid input-output model applied to fossil- and bio-based plastics. MethodsX, v. 8, 101525. https://doi.org/10.1016/j.mex.2021.101525
29. Jander, W. 2022. Advancing bioeconomy monitorings: A case for considering bioplastics. Sustainable Production and Consumption, v. 30, pp. 255-268. https://doi.org/10.1016/j.spc.2021.11.033
30. Jander, W., Grundmann, P. 2019. Monitoring the transition towards a bioeconomy: A general framework and a specific indicator. Journal of Cleaner Production, v. 236, 117564. https://doi.org/10.1016/j.jclepro.2019.07.039
31. Jander, Wiebke, Sven Wydra, Johann Wackerbauer, Philipp Grundmann, and Stephan Piotrowski. 2020. "Monitoring Bioeconomy Transitions with Economic–Environmental and Innovation Indicators: Addressing Data Gaps in the Short Term" Sustainability 12, no. 11: 4683. https://doi.org/10.3390/su12114683
32. Jasinevičius, Gediminas, Marcus Lindner, Pieter Johannes Verkerk, and Marius Aleinikovas. 2017. "Assessing Impacts of Wood Utilisation Scenarios for a Lithuanian Bioeconomy: Impacts on Carbon in Forests and Harvested Wood Products and on the Socio-Economic Performance of the Forest-Based Sector" Forests 8, no. 4: 133. https://doi.org/10.3390/f8040133
33. Jonsson, R. et al. 2021. Boosting the EU forest-based bioeconomy: Market, climate, and employment impacts. Technological Forescasting & Social Change, v. 163, 120478. https://doi.org/10.1016/j.techfore.2020.120478
34. Jurga, Piotr, Efstratios Loizou, and Stelios Rozakis. 2021. "Comparing Bioeconomy Potential at National vs. Regional Level Employing Input-Output Modeling" Energies 14, no. 6: 1714. https://doi.org/10.3390/en14061714
35. Kalogiannidis, S. et al. 2023. The Assessment of the Bioeconomy and Biomass Sectors in Central and Eastern European Countries. International Journal of Energy Economics and Policy, v. 13, n. 3, pp. 494-506. DOI: https://doi.org/10.32479/ijeep.14230
36. Kardung, Maximilian, Kutay Cingiz, Ortwin Costenoble, Roel Delahaye, Wim Heijman, Marko Lovrić, Myrna van Leeuwen, Robert M'Barek, Hans van Meijl, Stephan Piotrowski, and et al. 2021. "Development of the

Circular Bioeconomy: Drivers and Indicators" Sustainability 13, no. 1: 413. https://doi.org/10.3390/su13010413
37. Kuosmanen, T.; Kuosmanen, N.; El Meligi, A.; Ronzon, T.; Gurria Albusac, P.; Iost, S.; M'Barek, R. How Big is the Bioeconomy? Reflections from an economic perspective; Publications Office of the European Union: Luxembourg, 2020.
38. Lazorcakova, E., Dries, L., Peerlings, +J9:J10J. Pokrivcak, J. 2022. Potential of the bioeconomy in Visegrad countries: An input-output approach. Biomass and Bioenergy, v. 158, 106366. https://doi.org/10.1016/j.biombioe.2022.106366
39. Lee, D.-H. Bio-based economies in Asia: Economic analysis of development of bio-based industry in China, India, Japan, Korea, Malaysia and Taiwan. Int. J. Hydrogen Energy 2016, 41, 4333–4346. https://doi.org/10.1016/j.ijhydene.2015.10.048
40. Lehtonen, O., Okkonen, L. Regional socio-economic impacts of decentralised bioeconomy: a case of Suutela wooden village, Finland. Environ Dev Sustain 15, 245–256 (2013). https://doi.org/10.1007/s10668-012-9372-6
41. Lestan, Filip, Babu George, and Sajal Kabiraj. 2021. "Economic Performance and Composition of Nordic Bioeconomy Sectors (NBES)" Journal of Risk and Financial Management 14, no. 9: 418. https://doi.org/10.3390/jrfm14090418
42. Liobikiene, G. and Brizga, J. 2019. The challenges of bioeconomy implementation considering environmental aspects in the Baltic States: an input-output approach, in: International Conference Economic Science for Rural Development, Jelgava, May 2019, pp. 355–362, https://doi.org/10.22616/ESRD.2019.142.
43. Liobikiene, G. et al. 2020. The trends in bioeconomy development in the European Union: Exploiting capacity and productivity measures based on the land footprint approach. Land Use Policy, v. 91, 104375. https://doi.org/10.1016/j.landusepol.2019.104375
44. Loizou, E., Jurga, P., Rozakis, S. Faber, A. 2019. Assessing the Potentials of Bioeconomy Sectors in Poland Employing Input-Output Modeling. Sustainability, v. 11, n. 3, 594. https://doi.org/10.3390/su11030594
45. Ludwik, W., and Wicka, A. Bio-economy sector in Poland and its importance in the economy. Proceedings of the 2016 International Conference "ECONOMIC SCIENCE FOR RURAL DEVELOPMENT" No 41 .
46. M'Barek, R. et al. 2018. Getting (some) numbers right – derived economic indicators for the bioeconomy. EUR 29353 EN, Publications Office of the European Union, Luxembourg, 2018, ISBN 978-92-79-93907-5, doi:10.2760/2037, JRC113252.
47. Mainar-Causapé, A. J. et al. (2021). Constructing an open access economy-wide database for bioeconomy impact assessment in the European Union member states, Economic Systems Research, 33:2, 133-156, DOI: 10.1080/09535314.2020.1785848
48. Mainar-Causapè, A., 2017. Analysis of structural patterns in highly disaggregated bioeconomy sectors by EU Membre states using SAM/IO multipliers. JRC Tehchnical reports. European Commission-Joint research Centre. http://publications.jrc.ec.europa.eu/repository/bitstream/JRC106676/kj-na-28591-enn_pdf. Accessed 7.11.2018. pdf. Accessed 7.11.2018.
49. Mainar-Causapé, A.J. (2019). Análisis de los sectores de Bioeconomía a través de matrices de contabilidad social específicas (BioSAMs): el caso de España. Investigaciones Regionales - Journal of Regional Research, 2019/3 (45), 273-282.
50. Mainar-Causapé, A.J.; Philippidis, G. (Ed.), 2018. BioSAMs for the EU Member States. Constructing Social Accounting Matrices with a detailed disaggregation of the bio-economy, EUR 29235 EN, Publications Office of the European Union, Luxembourg, ISBN 978-92-79-85966-3, doi:10.2760/811691, PUBSY No. JRC111812.
51. Mikkelsen, E. Value added in the Norwegian Bioeconomy; NORUT Report 8/2017; Norut (Norut Northern Research Institute AS): Tromsø, Norway, 2017.
52. Morone, P. et al. 2022. Inter-connected challenges: an overview of bioeconomy in Europe. Environmental Research Letters, v. 17, n. 11, 114031. DOI 10.1088/1748-9326/ac9dac
53. Mõtte, M. et al., 2019. A systematic approach to exploring the role of primary sector in the development of Estonian bioeconomy. Agronomy Research, v. 17, n. 1, pp. 220-233. https://doi.org/10.15159/AR.19.068

54. MUNGARAY-MOCTEZUMA, A. B., Sylvia Monica PEREZ-NUÑEZ, Santos LOPEZLEYVA, 2015. Knowledge-Based Economy in Argentina, Costa Rica and Mexico: A Comparative Analysis from the Bio-Economy Perspective. Management Dynamics in the Knowledge Economy, v. 3. n. 2, pp. 213-236.
55. Nowak, Anna, Anna Kobiałka, and Artur Krukowski. 2021. "Significance of Agriculture for Bioeconomy in the Member States of the European Union" Sustainability 13, no. 16: 8709. https://doi.org/10.3390/su13168709
56. Nowicki P, Banse M, Bolck C, Bos H, Scott E. 2008. Biobased economy. State-of-the-art assessment. Rep., Agric. Econ. Res. Inst., The Hague, Neth.
57. O'Donoghue, C. et al. 2019., Measuring GHG emissions across the agri-food sector value chain: the development of a bioeconomy input-output model, Int. J. Food Syst. Dynam. 10 (1) (2019) 55–85, https://doi.org/10.18461/ijfsd.v10i1.04.
58. Pellerin W, Taylor DW. Measuring the biobased economy: a Canadian perspective. Ind Biotechnol 2008;4(4):363e6.
59. Perunová, M., Zimmermannová, J. 2022. Analysis of forestry employment within the bioeconomy labour market in the Czech Republic. Journal of Forest Science, v. 68, n. 10, pp. 385-394. https://doi.org/10.17221/84/2022-JFS
60. Philippidis, G., M'barek, R, & Ferrari, E. (2015). Drivers of the Bioeconomy in Europe towards 2030 - short overview of an exploratory, model-based assessment". European Commission, Joint Research Centre, Institute for Prospective Technological Studies, Spain.
61. Philippidis, G., SanjuánA. I., FerrariE., & M'barekR. (2014). Employing social accounting matrix multipliers to profile the bioeconomy in the EU member states: is there a structural pattern? Spanish Journal of Agricultural Research, 12(4), 913-926. https://doi.org/10.5424/sjar/2014124-6192
62. Philippidis, George, and Ana I. Sanjuán-López. 2018. "A Re-Examination of the Structural Diversity of Biobased Activities and Regions across the EU" Sustainability 10, no. 11: 4325. https://doi.org/10.3390/su10114325
63. Piotrowski, S.; Carus, M.; Carrez, D. European Bioeconomy in Figures 2008–2015; Nova-Institute for Ecology. and Innovation: Hürth, Germany, 2018; p. 16. Available online: http://biconsortium.eu/sites/biconsortium.eu/files/documents/Bioeconomy_data_2015_20150218.pdf
64. Piotrowski, S.; Carus, M.; Carrez, D. European Bioeconomy in Figures 2008–2016; nova-Institute for Ecology and Innovation: Hürth, Germany, 2019; p. 25. https://biconsortium.eu/file/1909/download?token=orOnanCb
65. Pokataiev, P. et al. 2023. The role of biotechnology in the development of the bioeconomy. Acta Innovations, v. 46, n. 2, pp. 18-33. https://doi.org/10.32933/ActaInnovations.46.2
66. Robert, N. et al. 2020. The EU Bioeconomy: Supporting an Employment Shift Downstream in theWood-Based Value Chains. Sustainability, v. 12, n. 3, 758. https://doi.org/10.3390/su12030758
67. Ronzon T, Santini F, M'Barek R. 2015. The bioeconomy in the European Union in numbers. Facts and figures on biomass, turnover and employment. Rep., Eur. Comm., Joint Res. Cent., Inst. Prospect. Tech. Stud., Spain Sevilla,
68. Ronzon, T., Iost, S. & Philippidis, G. Has the European Union entered a bioeconomy transition? Combining an output-based approach with a shift-share analysis. Environ Dev Sustain 24, 8195–8217 (2022). https://doi.org/10.1007/s10668-021-01780-8
69. Ronzon, T., Iost, S., Philippidis, G. 2022. An output-based measurement of EU bioeconomy services: Marrying statistics with policy insight. Structural Change and Economic Dynamics, v. 60, pp. 290-301. https://doi.org/10.1016/j.strueco.2021.10.005
70. Ronzon, T., M'Barek, R. 2018. Socioeconomic Indicators to Monitor the EU's Bioeconomy in Transition. Sustainability, v. 10, n. 6, 1745. https://doi.org/10.3390/su10061745
71. Ronzon, T., Piotrowski, S., M'Barek, R., & Carus, M. (2017). A systematic approach to understanding and quantifying the EU's bioeconomy. Bio-Based and Applied Economics, 6(1), 1–17. https://doi.org/10.13128/BAE-20567
72. Ronzon, T.; Piotrowski, S.; Tamosiunas, S.; Dammer, L.; Carus, M.; M'barek, R. Developments of Economic Growth and Employment in Bioeconomy Sectors across the EU. Sustainability 2020, 12, 4507.
73. Rosegrant MW, Ringler C, Zhu T, Tokgoz S, Bhandary P. 2013. Water and food in the bioeconomy: challenges and opportunities for development. Agric. Econ. 44(s1):139–50

74. Skorwider-Namiotko, J. (2015). LEVEL OF DEVELOPMENT OF BIOECONOMY IN POLAND ACCORDING TO THE REGIONAL APPROACH - MEASUREMENT TRIAL. Economic and Regional Studies, 8(1), 55-72.
75. T. Ronzon, M. Lusser, M. Klinkenberg (ed.), L. Landa, J. Sanchez Lopez (ed.), R. M'Barek, G. Hadjamu (ed.), A. Belward (ed.), A. Camia (ed.), J. Giuntoli, J. Cristobal, C. Parisi, E. Ferrari, L. Marelli, C. Torres de Matos, M. Gomez Barbero, E. Rodriguez Cerezo (2017). Bioeconomy Report 2016. JRC Scientific and Policy Report. EUR 28468 EN. https://op.europa.eu/en/publication-detail/-/publication/b3a3b800-4f18-11e7-a5ca-01aa75ed71a1/language-en
76. Talavyria, M. P., Lymar, V. V., & Baidala, V. V. (2017). Indicators for analysis of the bioeconomy in Ukraine. Економіка АПК, 3, 44–50.
77. van de Pas, J. 2015. The bio-economy: definitions and measurement, Wageningen University, The Netherlands. edepot.wur.nl)
78. van Meijl, H., Tsiropoulos, I., Bartelings, H., Hoefnagels, R., Smeets, E., Tabeau, A., & Faaij, A. (2018). On the macro-economic impact of bioenergy and biochemicals—Introducing advanced bioeconomy sectors into an economic modelling framework with a case study for the Netherlands. BIomass and Bioenergy, 108, 381–397. https ://doi.org/10.1016/j.biomb ioe.2017.10.040.
79. Vandermeulen V, Prins W, Nolte S, Van Huylenbroeck G. 2011. How to measure the size of a bio-based economy: evidence from Flanders. Biomass Bioenerg. 35:4368–75
80. Vlad, Ionela Mițuko, and Elena Toma. 2022. "The Assessment of the Bioeconomy and Biomass Sectors in Central and Eastern European Countries" Agronomy 12, no. 4: 880. https://doi.org/10.3390/agronomy12040880
81. Wen, Xuezhou, Daniel Quacoe, Dinah Quacoe, Kingsley Appiah, and Bertha Ada Danso. 2019. "Analysis on Bioeconomy's Contribution to GDP: Evidence from Japan" Sustainability 11, no. 3: 712. https://doi.org/10.3390/su11030712
82. Wydra, S. 2020. Measuring innovation in the bioeconomy – Conceptual discussion and empirical experiences. Technology in Society, v. 61, 101242. https://doi.org/10.1016/j.techsoc.2020.101242
83. Zargar, S. et al. 2022. The Application of Industrial Ecology Methods to Understand the Environmental and Economic Implications of the Forest Product Industries. Current Forestry Reports, v. 8, pp. 346-361. https://doi.org/10.1007/s40725-022-00174-x
84. Zimmermannová, J., Perunová, M. 2022. Bioeconomy labour market and its drivers in the Cazech Republic. Economics Management Innovation, v. 14, n. 1. 33-46.
85. Zúniga González, C. A., & Trejos, R. (2014). Medición de la contribución de la Bioeconomía: Caso Nicaragua. Universitas (León): Revista Científica De La UNAN León, 5(1), 59–82. https://doi.org/10.5377/universitas.v5i1.1479.

**Appendix B.** Code structure: concepts, methods, data, indicators, limitations, countries.

**Categories**
Codes

**Bioeconomy Concept**
BmBB
BsBB
BtBB

**Methods/Models**
A-LCA
AgroSAM
AHP
BioSAM
BTF
CGE
Cluster Analysis
Correlation Analysis
Cross-Entropy Method
Cross-sectional Analysis
Data Compilation and Estimates
Descriptive Statistics
Econometric Model
EEIO
Euclidean Distance
Experts Interview
Experts Judgement
Extended Hybrid IOM
GTAP Model
HEM
IMPLAN IO Model
Input- and Output-based Approach
Interpolation Estimates
IOM
LCA
Linear Regression Model
Literature Review
MacroSAM
MAGNET Model
MCDA
MRIO

Multiple Regression Analysis
Overview
Pairwise Comparisons
Questionnaire
RAS
SAM
Scenarios
Sensitivity Analysis
Structural Analysis
Variance Analysis

**Other Methods/Models**

ADF
Advanced LCA
AIC
ARCH Test
ARDL
BIO Model
Bottom-Up Approach
Breusch–Godfrey LM Test
BTS
Case Studies
Catch-up Rate
Causality Analysis
CAWI
CBM
CED
CES
CET
CH Rule
C-LCA
Cointegration Test
Comparative Analysis
Cross-sectional Survey
CUSUMQ Test
Decoupling Analysis
Desk Research
Diagnostic Tests
Dialectical Method
Distance-based Method
Dynamic Analysis
ECT
EEMRIO

EFISCEN
Eigenvector Method
FarSo Model
Flegg Location Quotient (FLQ)
Footprint Analysis
Forecasting
GAMS
GFN
GFTM
Ghosh Model
GINFORS Model
Granger Causality Test
GRIT Technique
GTEM
HAC Technique
Hellwig's Method
Historical-Logical Method
Hybrid IO-LCA Model
Impact Assessment Method
IMPACT Model
Intermediate Demand Shock
Jarque-Bera Test
Kaiser Criteria
KMO Test
Kruskal-Wallis
LandSHIFT Model
LCIA
Likert Scale
Linear Interpolation
LPES Model
Macroeconomic Modelling
Mann-Whitney Test
MARKAL-NL-UU
Matrix Augmentation Method
MFA
MicroSAM
Mixed Methods
Modified Regional IOM
Monte Carlo Simulation
Multivariate Analysis
PANTA RHEI Model
PCA
PP Test

Qualitative Approach
Quantitative Approach
Ramsey RESET Test
Rasmussen-Jones Method
Regional Energy Systems Model
Shift-share Analysis
SIMSIPSAM
SQL Technique
Survey
System Structural Method
Text Mining
Three Sectors Model
Time Series Analysis
Top-Down Approach
TOPSIS
ToSIA
Trend Analysis
Two Sectors Model
Value Chain Approach
Varimax
VEC Model
WALD Test
Ward Linkage
Web Scrapping

**Data Sources**

AgroSAM Data
BioSAM Data
CAPRI
CoCo Database
CPA
CZSO
DataM Database
Destatis
Diverse Private Data
Diverse Public Data
EAA
Ecoinvent
European Commission, JCR
Eurostat
EXIOBASE
Experts or Professionals
FAO

FAOSTAT
Forestry Employment
German Federal Statistical Office
GTAP Data
Inputs
IPCC
ISIC
LFS
MAGNET Database
NACE
NAICS
National Accounts
National IO Tables
Nova Institut
OECD
Primary Data
PRODCOM
Regional Accounts
SAM Database
SBS
Scientific Literature
Scopus
STECF
SUTs
Web of Science
WIOD
World Bank

**Other Data Sources**

BioMonitor
CAIT Climate Data Explorer
Central Bank of Hungary
Central Bank of Nicaragua
Central Statistical Bank
CICES
COFOG
COMEXT Database
Companies Information
CPC
CRF
CSO
Czech National Bank
EFISCEN Data

EIA
EPA
EU CSO
EU KLEMS
European Central Bank
European Patent Organization
European System of National and Regional Accounts
FADN
FEA
Finnish Forest Research Institute
Fraunhofer ISI
GEIH
GFTM Data
Global Footprint Network Database
GMFD
Google Scholar
GTAP-AEZ
GTAP-AGR
GTAP-E
Hungarian Central Statistical Office
IEA
IMAGE-TIME
IRENA
ISTAT
MARKAL-NL-UU Database
NAMEA
National Bank of Poland
ORBIS
PATSTAT
PKD
Regional I-O Tables
REGON
Research Gate
SAT-BBE
SEAI
Social Insurance Institution of Finland
SSS Ukraine
StatBank Norway
Statistical Office of the Slovak Republic
Statistics Canada
Statistics Estonia
Statistics Finland
Statistics Norway

Statistics Poland
Teagasc NFS
Technical Literature
US, BEA
WPINDEX - STN International

**Indicators**

Absorption Multiplier Effect
Agricultural Land
Backward Linkage
Bio-based Shares
Bioeconomy Share in Value Added
Biomass Use or Consumption
CO2 Emissions
Diffusion Multiplier Effect
Direct Effect
Employment
Employment Change
Employment Elasticity
Employment Multiplier
Employment Share
Exports
Forest Area
Forestry Biomass Production
Forward Linkage
Fossil and Mineral Resource Use
Fossil Resource Savings (FRS)
GDP ($)
GDP Growth Rate
GDP per capita
GHG Emissions
Gross Enrollment Ratio
Imports
Income Elasticity
Income Multiplier
Indirect Effect
Induced Effect
Innovation
Investment
Key Sectors
Labour Productivity
Land Footprint

Location Quotient (LQ)
Multipliers effects
Number of Jobs or Workers
Number of Publications
Operating Surplus and Mixed Income
Output Elasticity
Output Multipliers
Patent Applications
Product or Output Value
Production Volume
R&D Investments
Sales
SEIB
Share of renewable energy in electricity production or consumption
Subsidies
Substitution Fossil x Renewable
Substitution Share Indicator (SSI)
Total Effect
Turnover ($)
Turnover per capita
Value Added
Value Added x GDP
Wages and salaries
Water Consumption

**Other Indicators**

Access to Food
Acidification
Agricultural Biomass Production
Agro-biotechnology
Arable Land
Availability of Food
Average Annual Catch-up Growth Rate
Balassa Index
Biocapacity
Biochemical Share
Biodiversity
Bioeconomy Innovation Share
Bioeconomy-driving Policies
Biological Raw Material Consumption
Biomass contribution
Biomass Dry Matter
Biomass Self-sufficiency Rate

Bioproductive Land
Bio-related Higher Education Enrollment
Bottom-Up FRC
Capital Compensation
Carbon Storage
Carbon Storage Change
Certified Bio-based Products
CH4 Emissions
Circular Effects
Climate Change
Climate Change Adaptation
Climate Footprint
Comparative Advantage
Compensation of Employees
Composite Bioeconomy Sustainability Index
Computers per 1,000 inhabitants
Crop Land
Dataset
DEU
Development of Environmental Technologies
Direct Fossil Resource Intensity (dFRI)
DMI
Domestic Food Biomass Footprint
Domestic Material Consumption
Domestic Nonfood Biomass Footprint
Downstream Effect
Ecological Deficit or Reserve
Ecological Footprint
Economic Impact
Economic Performance
Education and Human Resources
Effects on Final Demand
Effects on LULC Changes
Effects on Trade Patterns
Eigenvector Multiplier
Electricity Production
Embedded Biological Raw Material
Employees with Advanced Education
Employment Reallocation Effect
Employment Within-sector Effect
Employment/Output Value
Energy Consumption
Energy Intensity

Environmental Investments
Export of Bio-based Products
Financing and government support
Fixed Assets
Food Supply Stability
Food Utilization
Fossil Chemical Share
Fossil Resource Consumption (FRC)
Freshwater ecotoxicity
Freshwater Eutrophication
GFCF
Global Innovation Index
Grass Land
Gross Margin
GWP
HEM Multiplier
High-Education Offer
Hours Worked
Human Development Index (HDI)
Human Toxicity - cancer effects
Human Toxicity - no cancer effects
ICT
Import of Bio-based Products
Imported Food Biomass Footprint
Imported Nonfood Biomass Footprint
Input Saving
Institution Regimen Index
International Co-publications
Ionising Radiation
Knowledge Economic Index (KEI)
Labor Compensation
Labor Productivity Growth
Land Cover
Land Intensity per Employees
Land Transformation
Literate Population
Manufactured Exports
Marine Eutrophication
Market Size
Market Value ($)
Material Footprint
Material Intensity per employees
Material Use Efficiency

N2O Emissions
Number of Bioeconomy Strategies
Number of employees per biotechnology company
Number of Research Groups
Number of sectoral economic entities per 10,000 inhabitants
Open Effects
Output Changes
Own Effects
Ozone Depletion
Particulate Matter
Percentage of recycled waste
Photochemical Ozone Formation
Prevalence of Undernurishment
Price Changes
Prices
Primary Biomass Production
Private Property Rights
Product Quantity
Production Change
Proportion of industrial innovative companies
Public Expenditure in Education
Public Funding Attractiveness Capacity
Rasmussen-Jones Multiplier
Recycling of biowaste
Recycling rate of wooden packing
Regulation
Relative advantage in environmental technologies
Renewable Energy Share
Residues Valorization
Resource Depletion
Resource sustainability
Resources productivity
Revenues
Revenues x GDP
Road Transport Fuel Share
RPA
Rural Population
Share of GFCF in the gross total expenditure
Share of land cultivated with certified organic farms
Share of new or significantly improved in the total sales
Skilled Labor
Socioeconomic impact
Students and Graduates

Sustainability of Bioeconomy Technology
Sustainable Resource Use
Synthetic Indicator of Bioeconomy Development
Telephones per 1,000 inhabitants
Terrestrial Eutrophication
Time
Top-Down FRC
Total Fossil Resource Intensity (tFRI)
Trade Balance
Turnover Growth Rate
Unemployment Rate
Upstream Effect
Users of Internet per 1,000 inhabitants
Value Added Growth
Value Added Multiplier
Value Added/Land-Use
Value Added/Material Consumption
Wage Costs
Water Resource Depletion
Workers Growth Rate
Public R&D x GDP
Turnover x GDP

**Geographic area**

All Countries
Antioquia
Argentina
Asia
Austria
Baltics
Belgium
Brazil
BRIC+
Bulgaria
Canada
CEE Countries
China
Colombia
Costa Rica
Croatia
Cyprus
Czech Republic
Denmark

Developed Countries
Developing Countries
East Asia and Pacific
Estonia
EU
EU Isles + Luxembourg
EU-10
EU-12
EU-13
EU-15
EU-2
EU-22
EU-24
EU-25
EU-26
EU-27
EU-28
EU-3
Europe and Central Asia
Europe Developed
Finland
Flanders
France
Germany
Greece
Hungary
Iceland
India
Indonesia
Ireland
Italian Regions
Italy
Japan
Korea
Latin America and Caribbean
Letonia
Lithuania
Lombardy Province
Lubelskie Region
Luxembourg
Main Eastern EU
Malaysia
Malta

Mediterranean EU
Mexico
Middle East and North Africa
NAFTA
NEMO
Nicaragua
Nordic Countries
North America
Northen & Central EU
Noruegian Counties
Norway
Poland
Portugal
Romania
ROW
Russia
Slovakia
Slovenia
South Asia
Spain
Sub-Saharan Africa
Suutela Village, Kontiolahti
Sweden
Taiwan
The Netherlands
U.S. States
Ukraine
United Kingdom
United States
Visegrad Countries
World